%% file: acl_latex-dyy.tex
\documentclass[11pt]{article}

\PassOptionsToPackage{table}{xcolor}
\usepackage[final]{acl}

\usepackage{times}
\usepackage{latexsym}

\usepackage[T1]{fontenc}

\usepackage[utf8]{inputenc}

\usepackage{microtype}

\usepackage{inconsolata}

\usepackage{graphicx}
\usepackage{amsmath}
\usepackage{amssymb}
\usepackage{booktabs}
\usepackage{url}
\usepackage{enumitem}

\usepackage{multirow}
\usepackage[table]{xcolor}
\usepackage{CJKutf8}
\newcommand{\zh}[1]{\begin{CJK*}{UTF8}{gbsn}#1\end{CJK*}}
\usepackage{tcolorbox}
\usepackage{CJKutf8}

\title{DuraS2ST: Chain-of-Thought and Reinforcement Learning for Duration-Aligned Speech-to-Speech Translation}

\author{
\begin{tabular}{c}
\textbf{
Yayue Deng\textsuperscript{1,2}\thanks{Work done during an internship at Microsoft},
Dingdong Wang\textsuperscript{1},
Yuxuan Hu\textsuperscript{2},
Jinyu Li\textsuperscript{2},
Yanqing Liu\textsuperscript{2},
} \\
\textbf{
Yuanyuan Wang\textsuperscript{1},
Weidong Chen\textsuperscript{1},
Helen M. Meng\textsuperscript{1},
Shujie Liu\textsuperscript{2}\footnotemark[2],
Xixin Wu\textsuperscript{1}\thanks{\ \ Corresponding Author}
} \\
\normalfont\textsuperscript{1}The Chinese University of Hong Kong, China \\
\normalfont\textsuperscript{2}Microsoft Corporation \\
\end{tabular}
}
  
\begin{document}
\maketitle

\begin{abstract}
Speech-to-speech translation (S2ST) in time-sensitive applications such as video dubbing requires not only semantic fidelity and speaker preservation, but also strict duration consistency to avoid audio-visual misalignment. However, existing S2ST systems largely generate target speech without explicit temporal planning, making duration control an unresolved challenge. We introduce \textbf{DuraS2ST}, a duration-aligned reasoning framework that enables a single speech language model to first generate an explicit chain-of-thought (CoT) for planning target wording and phonetic length, and then synthesize the corresponding speech tokens. To support this paradigm, we construct \textbf{DuraSet-440K}, a high-quality duration-aligned CoT corpus for supervised initialization. We further optimize the model with multi-modal multi-dimensional reinforcement learning, using a Duration Margin Reward to balance translation quality and duration consistency, and Modality-Aware Reward Attribution to assign rewards to appropriate token spans. Experiments on CVSS-T show that DuraS2ST achieves a strong balance between translation quality and duration consistency, outperforming competitive open-source and commercial baselines. Project page: \url{https://github.com/Mia11939/DuraS2ST}.
\end{abstract}

\input{sections/introduction}

\input{sections/method_new}

\input{sections/duraset}

\input{sections/experiments}

\input{sections/conclusion}

\input{sections/limitations}

\input{sections/acknowledgement}



\bibliography{custom}

\clearpage
\appendix

\twocolumn[%
\begin{center}
  \LARGE \bfseries Supplementary Material
\end{center}
\vspace{1em}
]

\input{sup-sections/additional_experiments}

\input{sup-sections/duraset}

\input{sup-sections/implementation_details}

\input{sup-sections/eval_example}

\input{sup-sections/subjective_eval_interface}

\end{document}

%% file: sections/introduction.tex
\section{Introduction}
Speech-to-speech translation (S2ST) aims to translate speech across languages while preserving the speaker's acoustic characteristics. Recent progress in end-to-end discrete unit modeling and large-scale Speech Language Models (SLMs) has significantly improved semantic fidelity and speaker preservation~\citep{lee2022direct, huang2023transpeech, inaguma2023unity, rubenstein2023audiopalm, barrault2023seamless, dong2024polyvoice}. However, most existing systems generate target speech without explicit temporal control. In time-sensitive scenarios such as video dubbing and simultaneous interpretation, duration mismatch can cause audio-visual desynchronization, overlapping speech, and degraded user experience~\citep{wu2023videodubber}.

\begin{figure}[t]
    \centering
    \includegraphics[width=\columnwidth]{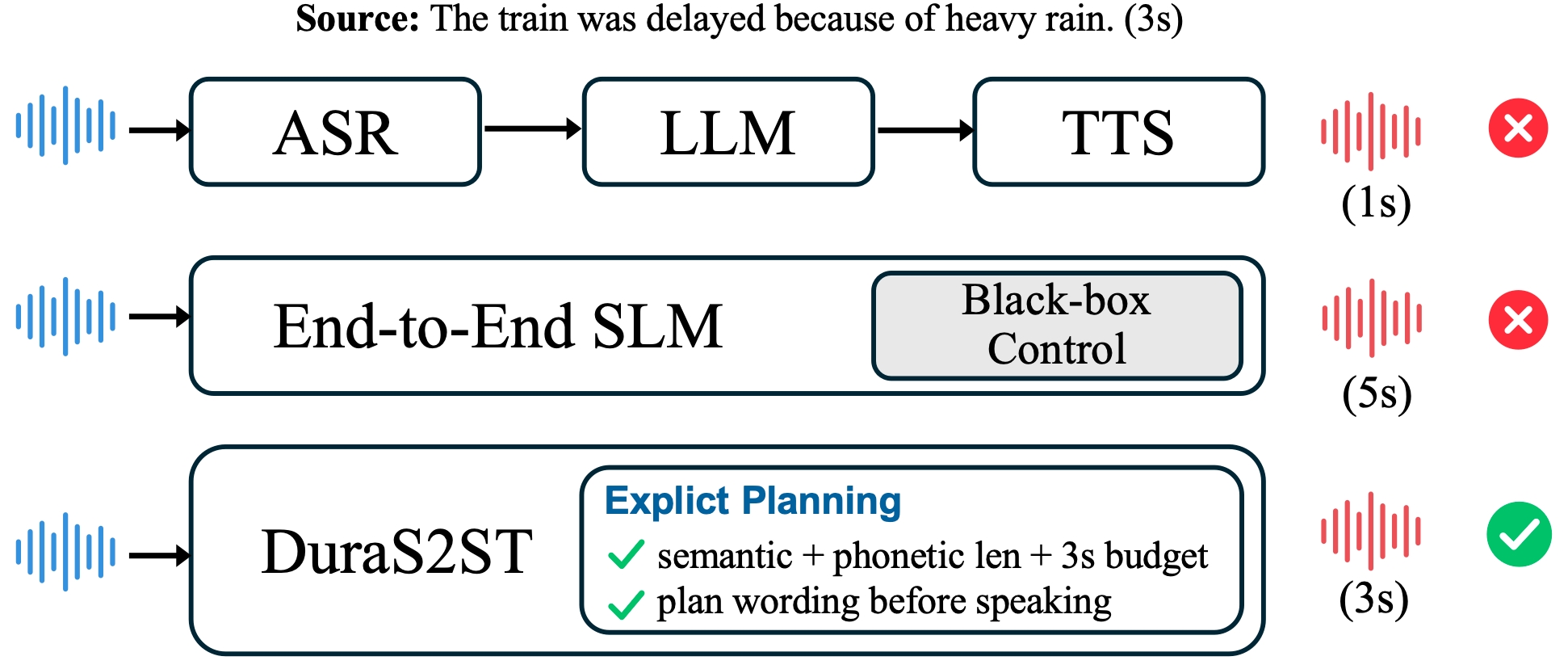}
    \caption{Existing systems suffer from duration mismatch: \zh{``你说得对''} can be translated as either ``You're right'' or ``I think you are absolutely right about that,'' leading to different speech durations. See Appendix~\ref{app:eval_example}.}
    \label{fig:teaser}
    \vspace{-7mm}
\end{figure}
Prior work on isochronous S2ST mainly follows two paradigms:cascaded and end-to-end systems. Cascaded systems~\citep{wahlsterverbmobil} often rely on post-hoc time-stretching, which may distort prosody and speaker timbre. End-to-end systems introduce speed tokens, length embeddings, or duration-conditioned generation~\citep{wu2023videodubber, le2024transvip, cheng2025seed}. While these methods improve temporal alignment, they typically treat duration as a low-level acoustic constraint: the decoder is encouraged to compress or expand speech to satisfy a target length. As shown in Figure~\ref{fig:teaser}, such black-box control lacks an explicit planning mechanism for balancing semantic content with phonetic length. In contrast, human dubbing often involves deliberate paraphrasing, where speakers adjust wording before articulation to fit the available time while preserving the core meaning.

Motivated by this observation, we propose \textbf{DuraS2ST}, a unified reasoning-based S2ST framework that formulates duration alignment as a planning problem rather than an acoustic post-processing problem. Built on a pre-trained SLM, DuraS2ST first generates an explicit reasoning trajectory that plans the target content and its phonetic footprint, and then produces the final interleaved text-acoustic sequence in a single autoregressive process. This allows semantic planning to directly guide subsequent speech generation.

DuraS2ST is trained in two stages. First, we perform supervised fine-tuning on a duration-aligned reasoning corpus that we construct, named \textbf{DuraSet-440K}. The model is supervised to plan, inside its chain-of-thought, both the target semantic content and its phonetic length before producing any acoustic token. This forces the model to actively balance semantic density against phonetic footprint at the symbolic level, replacing the black-box length control of speed tokens and length embeddings with explicit reasoning that drives subsequent acoustic synthesis. Second, we adopt Group Relative Policy Optimization \citep{shao2024deepseekmath} for our multi-modal generation task with a multi-dimensional reward design. We propose the \textbf{Duration Margin Reward} (DMR) to balance translation quality and duration consistency, and the \textbf{Modality-Aware Reward Attribution} (MARA) to shield the textual chain-of-thought from cross-modality reward contamination.
Our contributions are summarized as follows:
\begin{itemize}[leftmargin=*, topsep=2pt, itemsep=2pt, parsep=0pt]
    \item We propose \textbf{DuraS2ST}, the first unified framework that reformulates duration-aligned S2ST as a reasoning problem rather than an acoustic post-processing one. To support this paradigm, we further release \textbf{DuraSet-440K}, a large-scale duration-aligned reasoning corpus.

    \item We introduce a multimodal RL optimization strategy for duration-aligned S2ST. Specifically, \textbf{DMR} provides a smooth soft constraint for duration consistency, while \textbf{MARA} attributes reward signals to modality-specific token spans, preventing cross-modality reward contamination.

    \item Experiments on CVSS-T show that DuraS2ST achieves a strong balance among translation quality and duration consistency, outperforming competitive open-source and commercial baselines. Notably, we demonstrate that DuraS2ST achieves substantial gains even in a zero-shot RL setting.
\end{itemize}

%% file: sections/method_new.tex
\section{DuraS2ST}
\label{sec:method}
\begin{figure*}[t]
  \centering
  \includegraphics[width=\linewidth]{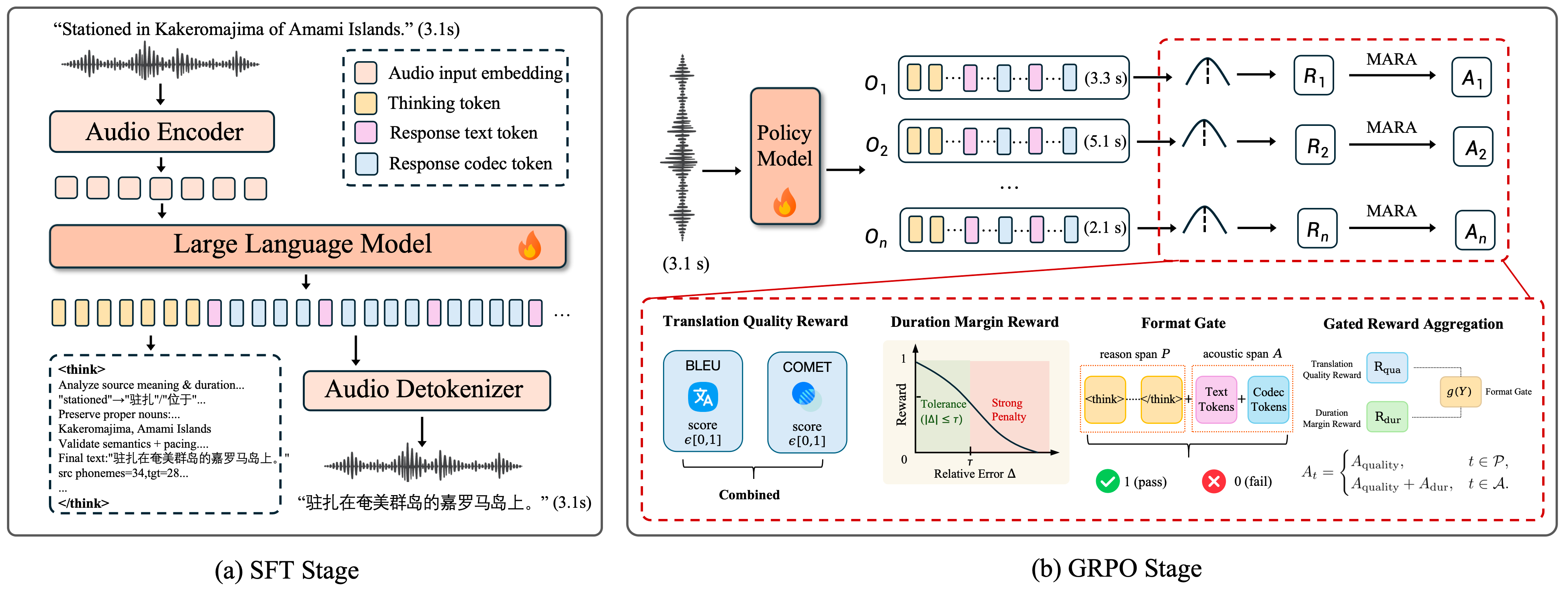}
  \caption{\textbf{Overview of the DuraS2ST framework.} DuraS2ST first generates a CoT rationale ($\mathbf{y}_{\mathrm{CoT}}$) for explicit duration planning, then synthesizes target speech tokens ($\mathbf{y}_{\mathrm{TA4}}$). The model is trained via a two-phase paradigm: SFT on DuraSet-440K, followed by GRPO with a multi-dimensional reward design that integrates translation quality, duration (DMR), and format compliance, dynamically attributed via MARA.}
  \label{fig:framework}
  \vspace{-4mm}
\end{figure*}

\vspace{-2mm}
To effectively resolve the duration mismatch in Speech-to-Speech Translation (S2ST), we propose \textbf{DuraS2ST}, as shown in Figure~\ref{fig:framework}, a unified framework built upon a single pre-trained multimodal language model. The central strength of DuraS2ST is its ability to explicitly negotiate the balance between semantic fidelity and acoustic duration. Unlike conventional cascaded pipelines or standard modular methods that treat duration control as a disconnected post-processing step, our approach leverages the multimodal language model's inherent reasoning capacity to formulate a rationale prior to acoustic execution, dynamically adjusting translation vocabulary and explicitly planning the resulting phoneme-sequence lengths to satisfy the duration budget.

\subsection{Phase I: SFT Training}
\label{subsec:sft}
As the first step of our training paradigm, we perform an SFT procedure \citep{wei2022finetuned} to instruct the SLM, Step-Audio-2-mini-Think \citep{wu2025step}, to adhere to strict duration constraints and execute a reasoning process in S2ST tasks.

Leveraging our constructed DuraSet-440K corpus (Section~\ref{sec:dataset}), denoted as $\mathcal{D}_{\mathrm{dura}}$, we conduct SFT to harness the model's pre-trained capabilities. Specifically, the model is guided to explicitly balance semantic fidelity and duration constraints by selecting the optimal translation vocabulary under the premise of maintaining comparable lengths between the source and target speech. For a given training sample ($\mathbf{X}_{\mathrm{src}}$, $\mathbf{Y}$), with $\mathbf{X}_{\mathrm{src}}$ representing the source speech representation, the output sequence $\mathbf{Y}$ is structured as a concatenation ($\oplus$) of two explicit phases:
\begin{equation}
\label{eq:output_format}
\resizebox{0.88\linewidth}{!}{$
    \!\underbrace{\text{\texttt{<think>}},\; \mathbf{y}_{\mathrm{CoT}},\; \text{\texttt{</think>}}}_{\text{\scriptsize \textsf{Reasoning Phase}}} 
    \oplus 
    \underbrace{\text{\texttt{<tts\_start>}},\; \mathbf{y}_{\mathrm{TA4}},\; \text{\texttt{<tts\_end>}}}_{\text{\scriptsize \textsf{Response Phase}}}\!
$},
\end{equation}
where $\mathbf{y}_{\mathrm{CoT}}$ represents the reasoning trajectory in which the model could jointly optimize semantic correspondence and length alignment, determining appropriate target words and sentence structures to match the source speech duration.
Conditioned on the reasoning results, the model generates $\mathbf{y}_{\mathrm{TA4}}$, an interleaved TA4 sequence in which each translated text (T) token is paired with four acoustic (A) tokens within a unified stream, following \citet{wu2025step}.



During training, the model is optimized to minimize the negative log-likelihood of the entire formatted target sequence $\mathbf{Y}$ given the source speech. 
Let $\mathcal{D}_{\mathrm{dura}}=\{(\mathbf{X}^{(i)}_{\mathrm{src}}, \mathbf{Y}^{(i)})\}_{i=1}^{N}$ denote DuraSet-440K, where $\mathbf{Y}^{(i)}=(\mathbf{y}^{(i)}_{\mathrm{CoT}}, \mathbf{y}^{(i)}_{\mathrm{TA4}})$ contains the CoT rationale and TA4 speech tokens. 
The SFT objective is formulated as:
\begin{equation}
\begin{aligned}
    \theta^*
    = \arg\min_\theta
    \sum_{i=1}^{N}
    -\log P_\theta\big(
        \mathbf{Y}^{(i)}
        \mid
        \mathbf{X}^{(i)}_{\mathrm{src}}
    \big),
\end{aligned}
\end{equation}
where $\theta$ denotes the trainable parameters of the SLM. 
The objective trains the model to autoregressively generate the CoT rationale and final speech tokens in a unified sequence. 
This explicitly conditions the TA4 output $\mathbf{y}_{\mathrm{TA4}}$ on both the source speech representation $\mathbf{X}_{\mathrm{src}}$ and the generated rationale $\mathbf{y}_{\mathrm{CoT}}$, thereby linking semantic planning with acoustic generation before the multimodal reinforcement learning phase.

\vspace{-2mm}
\subsection{Phase II: GRPO Training}
\label{subsec:rl}

Although SFT provides a strong initialization for reasoning and cross-modal alignment, it cannot directly optimize non-differentiable sequence-level objectives such as duration consistency and translation quality. We therefore introduce a multimodal reinforcement learning phase based on Group Relative Policy Optimization (GRPO)~\citep{shao2024deepseekmath}. For each query $q$, we sample a group of $G$ rollouts $\{o_1, \dots, o_G\}$ from the old policy $\pi_{\theta_{\text{old}}}$ and optimize them with translation-quality, duration, and format signals. 

This RL stage faces two task-specific challenges. First, strict duration matching can over-penalize minor deviations, forcing the model to trade translation fidelity and acoustic naturalness for length compliance. Second, because each trajectory interleaves textual CoT and acoustic tokens, sequence-level duration penalties may contaminate credit assignment by incorrectly penalizing preceding reasoning tokens. To address these issues, we first introduce Duration Margin Reward (DMR) as a soft duration objective, and then apply Modality-Aware Reward Attribution (MARA) to assign reward signals to appropriate token spans. We further incorporate translation-quality and format signals to preserve semantic fidelity and structural validity during GRPO optimization. We next describe the multi-modal multi-dimensional reward design, which consists of translation-quality, duration, and format signals, followed by the MARA mechanism for modality-aware span-level credit assignment.
\subsubsection{Multi-Modal Multi-Dimensional Reward Design}
\label{subsubsec:reward_design}

We use three complementary signals for GRPO: a translation-quality reward, a duration reward, and a format gate. The translation-quality reward encourages semantic fidelity, the duration reward encourages length consistency, and the format gate ensures that only structurally valid outputs receive positive optimization signals.

\paragraph{Translation Quality Reward.}
Let $\mathbf{Y}$ denote the raw generated multimodal sequence, $\mathbf{y}_{\text{text}}$ the text translation extracted from the TA4 stream by removing acoustic tokens, $\mathbf{x}$ the source text, and $\mathbf{y}^*$ the reference translation.

The translation-quality reward combines reference-based BLEU and reference-free COMET\textsubscript{Kiwi}:
\begin{equation}
    \resizebox{0.8\columnwidth}{!}{$\displaystyle
        R_{\text{qua}}
        =
        \text{BLEU}(\mathbf{y}_{\text{text}}, \mathbf{y}^*)
        +
        \text{COMET}(\mathbf{x}, \mathbf{y}_{\text{text}})
    $},
\end{equation}
where both scores are normalized to $[0,1]$.

\paragraph{Duration Margin Reward.}
To address the trade-off between duration consistency and translation quality, we model duration alignment as a bounded feasible region rather than a hard regression target. We introduce the Duration Margin Reward (DMR), a non-linear soft constraint that rewards outputs close to the target duration while tolerating minor deviations.

Let $L_{\text{pred}}$ and $L_{\text{tgt}}$ denote the predicted and target sequence lengths in acoustic tokens. We first compute the relative duration gap
$\delta = |L_{\text{pred}} - L_{\text{tgt}}| / L_{\text{tgt}}$,
and map it to a signed margin
$m = (\tau - \delta) / s$, where $\tau$ is the acceptable relative error bound and $s$ controls the smoothing scale. The duration reward is then defined as:
\begin{equation}
\resizebox{0.88\columnwidth}{!}{$\displaystyle
    R_{\text{dur}} = \frac{\sigma\big(\kappa \cdot \text{clip}(m, -B, B)\big) - \sigma(-\kappa B)}{\sigma(\kappa B) - \sigma(-\kappa B)} \in [0, 1]
$},
\end{equation}
where $\sigma(\cdot)$ is the Log-Sigmoid function, $\kappa$ controls the reward sharpness, and $B$ bounds the margin range. The min--max normalization rescales the bounded output to $[0,1]$, providing GRPO with a stable reward signal.

DMR is monotonically decreasing with the duration gap $\delta$: outputs within the acceptable region receive high rewards, while large deviations are smoothly suppressed. This design avoids over-penalizing minor duration differences, reduces sensitivity to outliers, and concentrates the learning signal around recoverable duration mismatches, thereby encouraging duration consistency without sacrificing translation fidelity or acoustic naturalness.

\paragraph{Format-Gated Aggregation.}
To ensure structural validity, we further define a binary format gate $g(\mathbf{Y}) \in \{0,1\}$, which equals $1$ only when $\mathbf{Y}$ follows the required ordered control tags and terminates correctly. Instead of using format compliance as an additive reward, we use it to gate the raw quality and duration rewards before MARA:
\begin{equation}
\label{eq:gated_reward}
\resizebox{0.88\columnwidth}{!}{$\displaystyle
    \bigl(r^{(i)}_{\text{qua}},\, r^{(i)}_{\text{dur}}\bigr)
    =
    g(\mathbf{Y}^{(i)})
    \cdot
    \bigl(R^{(i)}_{\text{qua}},\, R^{(i)}_{\text{dur}}\bigr).
$}
\end{equation}
Thus, malformed rollouts receive zero credit for both quality and duration, preventing reward hacking through format violations. The gated rewards $r^{(i)}_{\text{qua}}$ and $r^{(i)}_{\text{dur}}$ are then used by MARA for group normalization and span-level attribution.

\subsubsection{Modality-Aware Reward Attribution}
\label{subsubsec:mara}

Standard GRPO collapses all rewards into a single scalar advantage and broadcasts it to every generated token. This is suboptimal for our multimodal trajectories, where textual CoT tokens and acoustic tokens are governed by different objectives. In particular, applying the speech-specific duration reward to preceding reasoning tokens may penalize correct semantic planning for downstream acoustic length errors. We refer to this issue as Cross-Modality Reward Contamination.

To address this, we propose Modality-Aware Reward Attribution (MARA), which computes group-relative advantages separately for each reward component and assigns them only to the tokens they directly supervise. Given a group of $G$ rollouts $\{\mathbf{Y}^{(i)}\}_{i=1}^{G}$, we split each output into a reasoning span $\mathcal{P}$ 
and an acoustic span $\mathcal{A}$. For each reward component $k \in \{\text{qua},\, \text{dur}\}$, where qua stands for translation quality and dur for duration, we normalize rewards within the group:
\begin{equation}
\label{eq:mara_norm}
    \hat A^{(i)}_k
    =
    \frac{r^{(i)}_k - \mu_k}{\sigma_k + \varepsilon},
    \quad
    \mu_k
    =
    \tfrac{1}{G}\textstyle\sum_{j} r^{(j)}_k,
\end{equation}
where $\sigma_k$ is the within-group standard deviation and $\varepsilon$ is a small constant.

The span-level advantage is then assigned as:
\begin{equation}
\label{eq:mara_token}
    \hat A^{(i)}_t =
    \begin{cases}
        \hat A^{(i)}_{\text{qua}}, & t \in \mathcal{P}, \\
        \hat A^{(i)}_{\text{qua}} + \hat A^{(i)}_{\text{dur}}, & t \in \mathcal{A}.
    \end{cases}
\end{equation}
Here, $\hat A^{(i)}_{\text{qua}}$ is applied to both spans ($\mathcal{P}$ and $\mathcal{A}$), as it reflects the translation quality of both text and speech, while the duration advantage $\hat A^{(i)}_{\text{dur}}$ is applied only to the acoustic span $\mathcal{A}$, where length is manifested. MARA therefore preserves useful semantic feedback while preventing speech-specific penalties from contaminating textual reasoning.

%% file: sections/duraset.tex
\section{DuraSet-440K}
\label{sec:dataset}
\begin{figure*}[t]
  \centering
  \includegraphics[width=\linewidth]{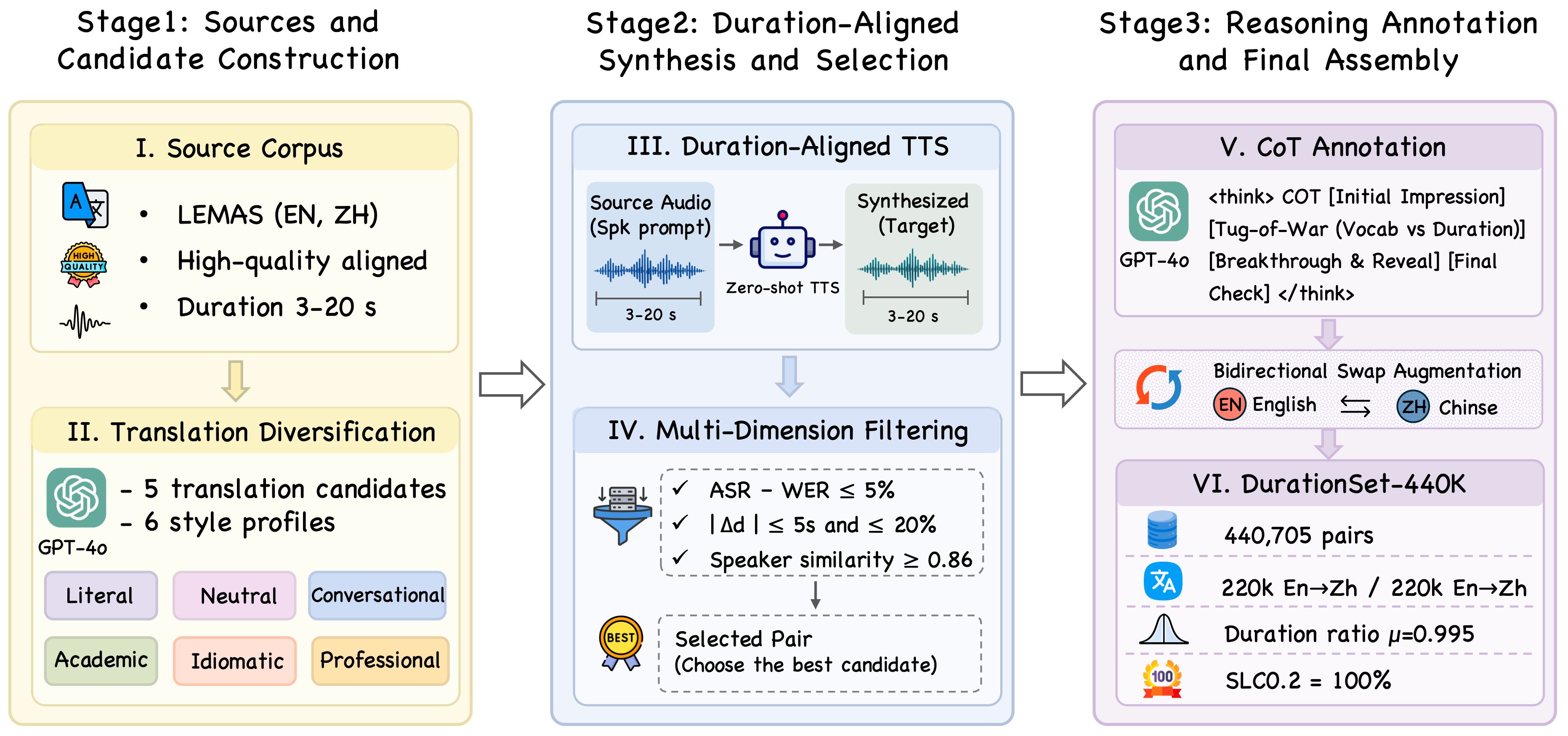}
  \caption{\textbf{Overview of the DuraSet-440K construction pipeline.}}
  \label{fig:pipeline}
  \vspace{-4mm}
\end{figure*}

To mitigate duration mismatch in speech-to-speech translation, we construct \textbf{DuraSet-440K}, a large-scale duration-aligned bilingual S2ST corpus containing 440{,}705 parallel utterances and roughly 1{,}000 hours of speech. It is designed to provide duration-aligned target speech that remains translation-faithful and speaker-consistent. We build the corpus from the English and Chinese subsets of LEMAS~\citep{zhao2026lemas}, first selecting high-quality utterances with suitable durations as source speech. Instead of relying on post-hoc time-stretching, we use duration-controllable synthesis to generate target speech that naturally aligns with the source duration. Figure~\ref{fig:pipeline} summarizes the overall construction pipeline, with detailed statistics and filtering settings provided in Appendix~\ref{sec:appendix-dataset}.

The pipeline contains three main stages. First, in \emph{Source and Candidate Construction}, we select English and Chinese utterances from LEMAS and filter them by quality and duration. Each retained source utterance is transcribed, and an LLM is prompted to generate multiple translation candidates with diverse styles and phonetic lengths. Second, in \emph{Duration-Aligned Synthesis and Selection}, each candidate is synthesized by a zero-shot duration-controllable TTS model conditioned on the source audio as a timbre prompt, and the best target pair is selected based on transcription accuracy, duration alignment, and speaker similarity. Third, in \emph{Reasoning Annotation and Final Assembly}, the retained pair is annotated with a phoneme-grounded reasoning trajectory and augmented by bidirectional source-target swapping, yielding the final duration-aligned bilingual S2ST corpus.


%% file: sections/experiments.tex
\section{Experiments}
\label{sec:experiments}

\subsection{Training Setup}
\label{sec:training_setup}

We build DuraS2ST on top of \textbf{Step-Audio-2-mini-Think}~\citep{wu2025step}\footnote{\url{https://huggingface.co/stepfun-ai/Step-Audio-2-mini-Think}}, a speech language model that natively interleaves text and discrete acoustic tokens. This makes it well suited for the structured CoT-to-speech format defined in Section~\ref{subsec:sft}, without requiring architectural modifications. Training proceeds in two phases.

\textbf{Phase I: SFT.}
We perform full-parameter fine-tuning on DuraSet-440K for $4{,}000$ steps using AdamW with a learning rate of $1\mathrm{e}{-5}$ and a global batch size of $128$. Each training trajectory follows the CoT-then-speech format in Eq.~\ref{eq:output_format}.

\textbf{Phase II: GRPO.}
Starting from the SFT checkpoint, we further optimize the model with GRPO for $1{,}000$ steps, using a learning rate of $1.5\mathrm{e}{-5}$ and a KL coefficient of $0.001$. To reduce training cost, we attach LoRA~\citep{hu2022lora} adapters to all linear projections and use vLLM~\citep{kwon2023efficient} for efficient rollout generation.

The reward combines translation quality and duration consistency, with malformed outputs filtered by the format gate. Specifically, the quality reward combines BLEU and COMET\textsubscript{Kiwi}, while the duration reward uses the proposed margin-based formulation with tolerance $\tau{=}5\%$. These rewards are converted into independent group-relative advantages through MARA, and the format gate zeroes both rewards for malformed rollouts before attribution. We also apply dynamic sampling~\citep{yu2026dapo} for training stability. Full hyperparameters and reward implementation details are provided in Appendix~\ref{app:training} and Appendix~\ref{app:reward}.

\subsection{Evaluation Setup}
\label{sec:evaluation_setup}

\subsubsection{Test Data}
We evaluate on the Chinese (ZH) and English (EN) test splits of \textbf{CVSS-T}~\citep{jia2022cvss}, comprising 4,897 utterance pairs, totaling 8.2 hours for Chinese and 6.3 hours for English. Since CVSS-T is disjoint from DuraSet-440K, it serves as an out-of-domain benchmark for evaluating translation quality and duration alignment.

\subsubsection{Metrics}
We evaluate S2ST performance from three aspects: translation quality, voice/prosody preservation, and duration consistency. Evaluator checkpoints and implementation details are provided in Appendix~\ref{app:metric}.

\textbf{Translation Quality.}
We report \textbf{Text-BLEU} on the intermediate text output and \textbf{Speech-BLEU} on the ASR transcription of the generated speech, both computed with SacreBLEU~\citep{post2018sacrebleu}. We also report reference-based \textbf{COMET}~\citep{rei2020comet} and reference-free \textbf{COMET\textsubscript{Kiwi}}~\citep{rei2022cometkiwi}.

\textbf{Voice and Prosody.}
We report \textbf{A.PCP}~\citep{barrault2023seamless} to measure prosodic consistency, and \textbf{SECS} to measure speaker similarity using WavLM~\citep{chen2022wavlm} embeddings between the source prompt and generated speech.

\textbf{Duration Consistency.}
We report \textbf{SLC-0.2} and \textbf{SLC-0.4}~\citep{wu2023videodubber}, which measure the fraction of samples whose target-to-source duration ratio falls within $\pm20\%$ and $\pm40\%$, respectively. To capture fine-grained duration errors, we further report \textbf{MADE}, the mean absolute duration error in seconds, and \textbf{MRDE}, the mean relative duration error.

\input{tables/main_results_zh2en}
\input{tables/main_results_en2zh}

\subsubsection{Baselines}
We compare DuraS2ST with both open-source and commercial S2ST systems:
\begin{itemize}[leftmargin=*, topsep=2pt, itemsep=1pt]
    \item \textbf{UniSS}~\citep{cheng2025uniss}: a strong Chinese-English S2ST system with unified text-speech modeling and structured CoT prompting. We report both Quality and Performance modes, denoted as \textbf{UniSS\,(Q)} and \textbf{UniSS\,(P)}.
    \item \textbf{Qwen2.5-Omni-7B}~\citep{xu2025qwen25omni} and \textbf{Kimi-Audio}~\citep{ding2025kimi}: recent audio/omni-modal MLLMs with speech-to-speech generation capability.
    \item \textbf{Step-Audio-2-mini} and \textbf{Step-Audio-2-mini-Think}~\citep{wu2025step}: the non-reasoning and reasoning variants of our backbone model.
    \item \textbf{SeamlessM4T}~\citep{barrault2023seamlessm4t}: Meta's multilingual S2ST system, including Medium, Large, and v2-Large variants.
    \item \textbf{GPT-4o}: a closed-source commercial S2ST reference.
\end{itemize}
Checkpoint details and prompt templates for all baselines are provided in Appendix~\ref{app:baseline}.

We further evaluate four variants based on our backbone to isolate the effects of SFT, CoT supervision, and GRPO:
\begin{itemize}[leftmargin=*, topsep=2pt, itemsep=1pt]
    \item \textbf{DuraS2ST-Baseline}: the original Step-Audio-2-mini-Think backbone without additional training.
    \item \textbf{DuraS2ST-Instruct}: SFT on DuraSet-440K without the \texttt{<think>} CoT.
    \item \textbf{DuraS2ST-Think}: SFT on DuraSet-440K with the \texttt{<think>} CoT.
    \item \textbf{DuraS2ST-Think-RL}: GRPO training initialized from DuraS2ST-Think.
\end{itemize}

\subsubsection{Inference Configuration}


We serve DuraS2ST with vLLM~\citep{kwon2023efficient} using the official Step-Audio-2 sampling configuration: temperature $0.7$, top-$p{=}0.9$, frequency penalty $0$, and repetition penalty $1.05$. The generated acoustic tokens are converted to waveforms using Step-Audio-2's \texttt{token2wav} module~\citep{wu2025step}. All DuraS2ST variants use the same Step-Audio-2 prompt template as in data construction, with baseline-specific prompts listed in Appendix~\ref{app:baseline}.

\subsection{Main Results}
\label{sec:main_results}

Tables~\ref{tab:en2zh} and~\ref{tab:zh2en} report the main results on CVSS-T. Overall, DuraS2ST achieves the strongest balance across translation quality, acoustic preservation, and duration consistency. More importantly, the stepwise improvements from Baseline to Instruct, Think, and Think-RL validate the effectiveness of duration-aligned supervision, explicit CoT planning, and reinforcement learning.
\input{figures/zero_shot_grpo_curves}
\textbf{Translation quality.}
On EN-ZH, DuraS2ST-Think-RL improves AVG\textsubscript{BLEU} from 27.25 to 30.95 and achieves the best open-source COMET\textsubscript{Kiwi}. On ZH-EN, it obtains the best overall translation performance among all systems, reaching the highest Text-BLEU, Speech-BLEU, COMET, AVG\textsubscript{BLEU}, and AVG\textsubscript{COMET}. This suggests that our DuraS2ST framework can achieve translation quality comparable to or even better than strong models without explicit duration control, while delivering substantially stronger duration consistency. Qwen2.5-Omni occasionally appends assistant-like speech in direct S2ST generation, which lowers both translation and duration scores.

\textbf{Duration consistency.}
DuraS2ST-Think-RL brings substantial improvements in duration alignment. Compared with the backbone, it raises SLC-0.2 from 0.487 to 0.997 on EN-ZH and from 0.305 to 0.934 on ZH-EN, while reducing MRDE to the lowest values in both directions. It achieves the best results on most duration metrics and remains highly competitive on the others, showing the effectiveness of our duration-aware training in improving source-target synchronization.

\textbf{Effect of CoT planning.}
The comparison between DuraS2ST-Instruct and DuraS2ST-Think highlights the role of explicit reasoning supervision. Both models are trained on the same DuraSet-440K corpus, but only Think exposes the \texttt{<think>} trajectory before acoustic generation. Introducing this reasoning stage leads to clear gains in duration alignment, reducing MADE from 0.63s to 0.18s on EN-ZH and from 1.31s to 0.48s on ZH-EN, with similar reductions in MRDE. These results suggest that structured CoT mainly benefits duration planning by providing a semantic-phonetic intermediate step before acoustic token generation. While its effect on translation quality varies across directions, the subsequent RL stage further improves translation performance and duration consistency, showing the complementary roles of CoT-based planning and reward-based optimization.

\textbf{Effect of GRPO.}
The transition from DuraS2ST-Think to DuraS2ST-Think-RL demonstrates the benefit of reward-based optimization. GRPO consistently improves translation metrics in both directions, especially on ZH-EN, where AVG\textsubscript{BLEU} increases from 24.88 to 27.30. Meanwhile, duration consistency is largely preserved or further improved, with MRDE reduced from 0.039 to 0.038 on EN-ZH and from 0.079 to 0.068 on ZH-EN. These results show that the RL stage enhances semantic fidelity under duration-aligned generation, while DMR and MARA provide effective duration-aware reward attribution.

\textbf{Acoustic preservation.}
DuraS2ST maintains strong prosody and speaker similarity while enforcing duration consistency. On EN-ZH, the trained DuraS2ST variants achieve the best A.PCP, and Think-RL obtains the highest SECS in both directions. This is important because duration control via post-hoc time-stretching can distort speaker identity or naturalness. In contrast, DuraS2ST learns to satisfy duration constraints during generation, preserving acoustic characteristics while producing synchronized speech.
\input{tables/mos_results}

\vspace{-2mm}
\subsection{Subjective Evaluation}
\label{sec:mos}

We conduct a human evaluation with six raters on $160$ anonymized CVSS-T samples, covering eight systems and both translation directions. Each sample is presented in a blinded and randomized order, and rated on a $1$--$5$ Likert scale for translation adequacy, speech naturalness, and speaker similarity.

Table~\ref{tab:mos} shows that the Think and Think-RL variants are consistently preferred or remain highly competitive in subjective evaluation. DuraS2ST-Think-RL achieves the best overall performance on ZH-EN and remains competitive on EN-ZH. Together with the objective duration metrics, these results indicate that DuraS2ST improves synchronization without degrading speech quality.

\input{tables/zero_grpo_duration}

\subsection{Analysis}
\label{sec:analysis}

We examine whether the proposed reward design can guide duration-aware optimization without supervised duration-aligned training. We apply GRPO directly to the Step-Audio-2-mini-Think backbone without SFT on DuraSet-440K, denoted as Zero-GRPO.

Table~\ref{tab:zero_grpo_duration} shows that Zero-GRPO reduces duration errors over the backbone, particularly in MADE and MRDE. When the duration reward is removed, the gains become weaker, confirming that the Duration Margin Reward is the main driver of duration alignment. Figure~\ref{fig:zero_shot_grpo} further shows that GRPO improves the duration reward and aggregated reward during training. Together, these results indicate that duration consistency can be optimized as a sequence-level objective, even without supervised duration-aligned trajectories.

%% file: tables/main_results_zh2en.tex
\begin{table*}[t]
\centering
\setlength{\tabcolsep}{1.0mm}
\renewcommand{\arraystretch}{1.15}
\caption{Main results on CVSS-T (\textbf{ZH-EN}). AVG\textsubscript{BLEU} and AVG\textsubscript{COMET} are means of (Text-BLEU, Speech-BLEU) and (COMET, COMET\textsubscript{Kiwi}). \textbf{Best} / \underline{2nd--3rd} scores are highlighted. ``--'' indicates unavailable values.}
\label{tab:zh2en}
\small
\resizebox{\textwidth}{!}{%
\begin{tabular}{l cccccc cc cccc}
\toprule
\textbf{Model}
 & \multicolumn{6}{c}{\textbf{Translation Quality}}
 & \multicolumn{2}{c}{\textbf{Voice \& Prosody}}
 & \multicolumn{4}{c}{\textbf{Duration Consistency}} \\
\cmidrule(lr){2-7} \cmidrule(lr){8-9} \cmidrule(lr){10-13}
 & \textbf{Text-BLEU}$\uparrow$ & \textbf{Speech-BLEU}$\uparrow$
 & \textbf{COMET}$\uparrow$ & \textbf{COMET\textsubscript{Kiwi}}$\uparrow$
 & \textbf{AVG\textsubscript{BLEU}}$\uparrow$ & \textbf{AVG\textsubscript{COMET}}$\uparrow$
 & \textbf{A.PCP}$\uparrow$ & \textbf{SECS}$\uparrow$
 & \textbf{SLC-0.2}$\uparrow$ & \textbf{SLC-0.4}$\uparrow$ & \textbf{MADE\,(s)}$\downarrow$ & \textbf{MRDE}$\downarrow$ \\
\midrule
\multicolumn{13}{c}{\textit{General Speech Large Language Models}} \\
\midrule
GPT-4o                  & -- & 23.13 & 0.710 & 0.692 & -- & 0.701 & 2.49 & 0.018 & 0.442 & 0.836 & 1.59 & 0.246 \\
Qwen2.5-Omni            & 10.86 & 10.82 & 0.638 & 0.611 & 10.84 & 0.625 & 1.75 & 0.142 & 0.190 & 0.355 & 7.85 & 1.310 \\
Kimi-Audio              & 19.87 & 13.91 & 0.698 & 0.700 & 16.89 & 0.699 & 2.10 & 0.188 & 0.429 & 0.746 & 2.56 & 0.462 \\
Step-Audio-2-mini       & 25.42 & 20.58 & 0.752 & \textbf{0.743} & 23.01 & \underline{0.745} & 2.49 & \underline{0.511} & 0.312 & 0.681 & 1.98 & 0.348 \\
UniSS\,(Q)              & 25.86 & \underline{24.54} & 0.780 & 0.699 & \underline{25.20} & 0.740 & \underline{2.73} & 0.309 & 0.913 & \underline{0.983} & \underline{0.74} & 0.125 \\
UniSS\,(P)              & \underline{26.04} & 24.00 & 0.769 & 0.683 & 25.02 & 0.726 & \textbf{2.74} & 0.310 & 0.893 & \textbf{0.986} & \underline{0.74} & \underline{0.123} \\
\midrule
\multicolumn{13}{c}{\textit{Speech-to-Speech Translation Models}} \\
\midrule
SeamlessM4T-Medium      & 19.02 & 15.65 & 0.673 & 0.663 & 17.34 & 0.668 & 2.26 & 0.041 & 0.260 & 0.649 & 2.26 & 0.331 \\
SeamlessM4T-Large       & 22.20 & 19.52 & 0.682 & 0.677 & 20.86 & 0.680 & 2.24 & 0.039 & 0.365 & 0.768 & 1.97 & 0.294 \\
SeamlessM4T-v2-Large    & 23.57 & 22.77 & 0.704 & 0.703 & 23.17 & 0.704 & 2.10 & 0.041 & 0.312 & 0.713 & 2.06 & 0.310 \\
\midrule
\rowcolor{gray!15}
\textbf{DuraS2ST-Baseline} & 24.14 & 23.09 & 0.740 & \underline{0.740} & 23.62 & 0.740 & 2.49 & 0.453 & 0.305 & 0.680 & 2.22 & 0.396 \\
\rowcolor{gray!15}
\textbf{DuraS2ST-Instruct} & \underline{27.42} & \underline{24.81} & \underline{0.789} & 0.692 & \underline{26.12} & 0.741 & 2.65 & \underline{0.500} & \underline{0.926} & 0.978 & 1.31 & 0.170 \\
\rowcolor{gray!15}
\textbf{DuraS2ST-Think}    & 25.99 & 23.76 & \underline{0.790} & 0.699 & 24.88 & \underline{0.745} & \underline{2.71} & 0.499 & \underline{0.930} & 0.977 & \underline{0.48} & \underline{0.079} \\
\rowcolor{gray!15}
\textbf{DuraS2ST-Think-RL} & \textbf{28.71} & \textbf{25.88} & \textbf{0.794} & 0.700 & \textbf{27.30} & \textbf{0.747} & 2.70 & \textbf{0.518} & \textbf{0.934} & \underline{0.979} & \textbf{0.42} & \textbf{0.068} \\
\bottomrule
\end{tabular}}
\vspace{-2mm}
\end{table*}

%% file: tables/main_results_en2zh.tex
\begin{table*}[t]
\centering
\setlength{\tabcolsep}{1.0mm}
\renewcommand{\arraystretch}{1.15}
\caption{Main results on CVSS-T (\textbf{EN-ZH}). AVG\textsubscript{BLEU} and AVG\textsubscript{COMET} are means of (Text-BLEU, Speech-BLEU) and (COMET, COMET\textsubscript{Kiwi}). \textbf{Best} / \underline{2nd--3rd} scores are highlighted. ``--'' indicates unavailable values.}
\label{tab:en2zh}
\small
\resizebox{\textwidth}{!}{%
\begin{tabular}{l cccccc cc cccc}
\toprule
\textbf{Model}
 & \multicolumn{6}{c}{\textbf{Translation Quality}}
 & \multicolumn{2}{c}{\textbf{Voice \& Prosody}}
 & \multicolumn{4}{c}{\textbf{Duration Consistency}} \\
\cmidrule(lr){2-7} \cmidrule(lr){8-9} \cmidrule(lr){10-13}
 & \textbf{Text-BLEU}$\uparrow$ & \textbf{Speech-BLEU}$\uparrow$
 & \textbf{COMET}$\uparrow$ & \textbf{COMET\textsubscript{Kiwi}}$\uparrow$
 & \textbf{AVG\textsubscript{BLEU}}$\uparrow$ & \textbf{AVG\textsubscript{COMET}}$\uparrow$
 & \textbf{A.PCP}$\uparrow$ & \textbf{SECS}$\uparrow$
 & \textbf{SLC-0.2}$\uparrow$ & \textbf{SLC-0.4}$\uparrow$ & \textbf{MADE\,(s)}$\downarrow$ & \textbf{MRDE}$\downarrow$ \\
\midrule
\multicolumn{13}{c}{\textit{General Speech Large Language Models}} \\
\midrule
GPT-4o                  & -- & \underline{31.46} & \textbf{0.845} & \underline{0.780} & -- & \textbf{0.812} & 2.59 & 0.051 & 0.565 & 0.858 & 1.09 & 0.253 \\
Qwen2.5-Omni            & 15.72 & 15.47 & 0.706 & 0.693 & 15.60 & 0.700 & 2.17 & 0.082 & 0.272 & 0.414 & 4.78 & 1.259 \\
Kimi-Audio              & 27.65 & 19.13 & 0.820 & 0.768 & 23.39 & 0.794 & 2.28 & 0.096 & 0.571 & 0.830 & 3.21 & 0.923 \\
Step-Audio-2-mini       & \underline{32.47} & 26.66 & 0.812 & 0.762 & 29.57 & 0.785 & 2.75 & 0.330 & 0.471 & 0.852 & 1.58 & 0.423 \\
UniSS\,(Q)              & \textbf{33.03} & \textbf{32.16} & \underline{0.828} & 0.773 & \textbf{32.60} & 0.801 & 2.72 & 0.315 & 0.984 & 0.996 & 0.34 & 0.067 \\
UniSS\,(P)              & 30.69 & 29.93 & 0.817 & 0.766 & \underline{30.31} & 0.792 & 2.73 & 0.316 & \underline{0.990} & \underline{0.997} & 0.33 & 0.065 \\
\midrule
\multicolumn{13}{c}{\textit{Speech-to-Speech Translation Models}} \\
\midrule
SeamlessM4T-Medium      & 24.80 & 15.84 & 0.757 & 0.744 & 20.32 & 0.751 & 2.31 & 0.033 & 0.561 & 0.849 & 1.52 & 0.310 \\
SeamlessM4T-Large       & 27.33 & 19.57 & 0.775 & 0.763 & 23.45 & 0.769 & 2.29 & 0.030 & 0.618 & 0.846 & 1.52 & 0.275 \\
SeamlessM4T-v2-Large    & 26.66 & 24.36 & 0.767 & 0.742 & 25.51 & 0.755 & 2.39 & 0.055 & 0.660 & 0.913 & 0.89 & 0.183 \\
\midrule
\rowcolor{gray!15}
\textbf{DuraS2ST-Baseline} & 27.57 & 26.92 & 0.821 & 0.773 & 27.25 & 0.797 & 2.75 & 0.333 & 0.487 & 0.851 & 2.04 & 0.629 \\
\rowcolor{gray!15}
\textbf{DuraS2ST-Instruct} & 29.51 & 28.70 & 0.827 & \underline{0.780} & 29.11 & 0.804 & \textbf{2.79} & \underline{0.367} & 0.981 & 0.996 & 0.63 & 0.090 \\
\rowcolor{gray!15}
\textbf{DuraS2ST-Think}    & 30.46 & 29.39 & \underline{0.828} & \underline{0.782} & 29.93 & \underline{0.805} & \textbf{2.79} & \underline{0.370} & \underline{0.993} & \textbf{0.999} & \textbf{0.18} & \underline{0.039} \\
\rowcolor{gray!15}
\textbf{DuraS2ST-Think-RL} & \underline{31.47} & \underline{30.43} & \underline{0.834} & \textbf{0.783} & \underline{30.95} & \underline{0.808} & \textbf{2.79} & \textbf{0.379} & \textbf{0.997} & \textbf{0.999} & \underline{0.23} & \textbf{0.038} \\
\bottomrule
\end{tabular}}
\vspace{-2mm}
\end{table*}

%% file: figures/zero_shot_grpo_curves.tex
\begin{figure*}[t]
\centering
\begin{minipage}[t]{0.245\textwidth}
  \centering
  \includegraphics[width=\linewidth]{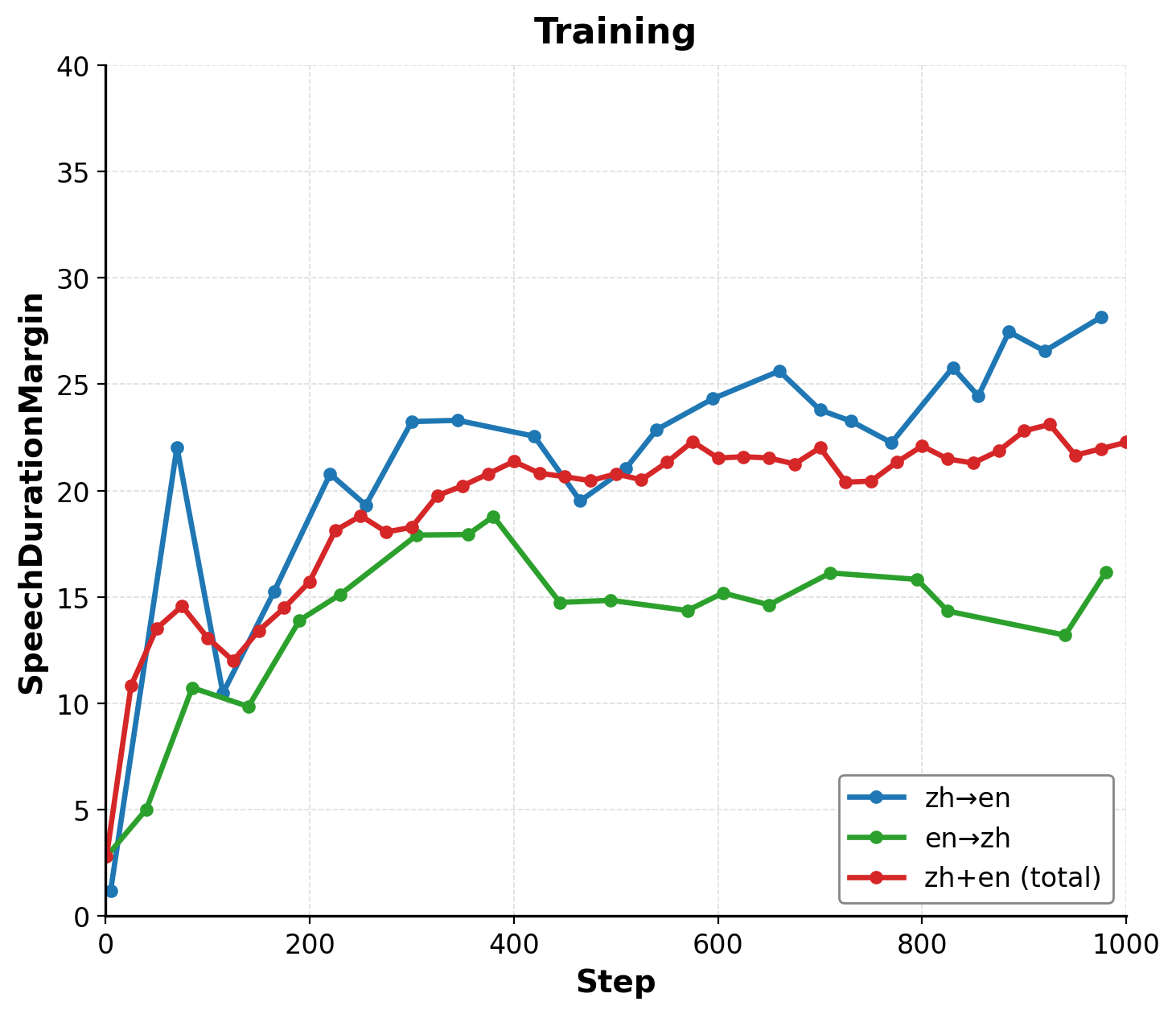}\\
  \small (a) Duration Margin Reward
\end{minipage}\hfill
\begin{minipage}[t]{0.245\textwidth}
  \centering
  \includegraphics[width=\linewidth]{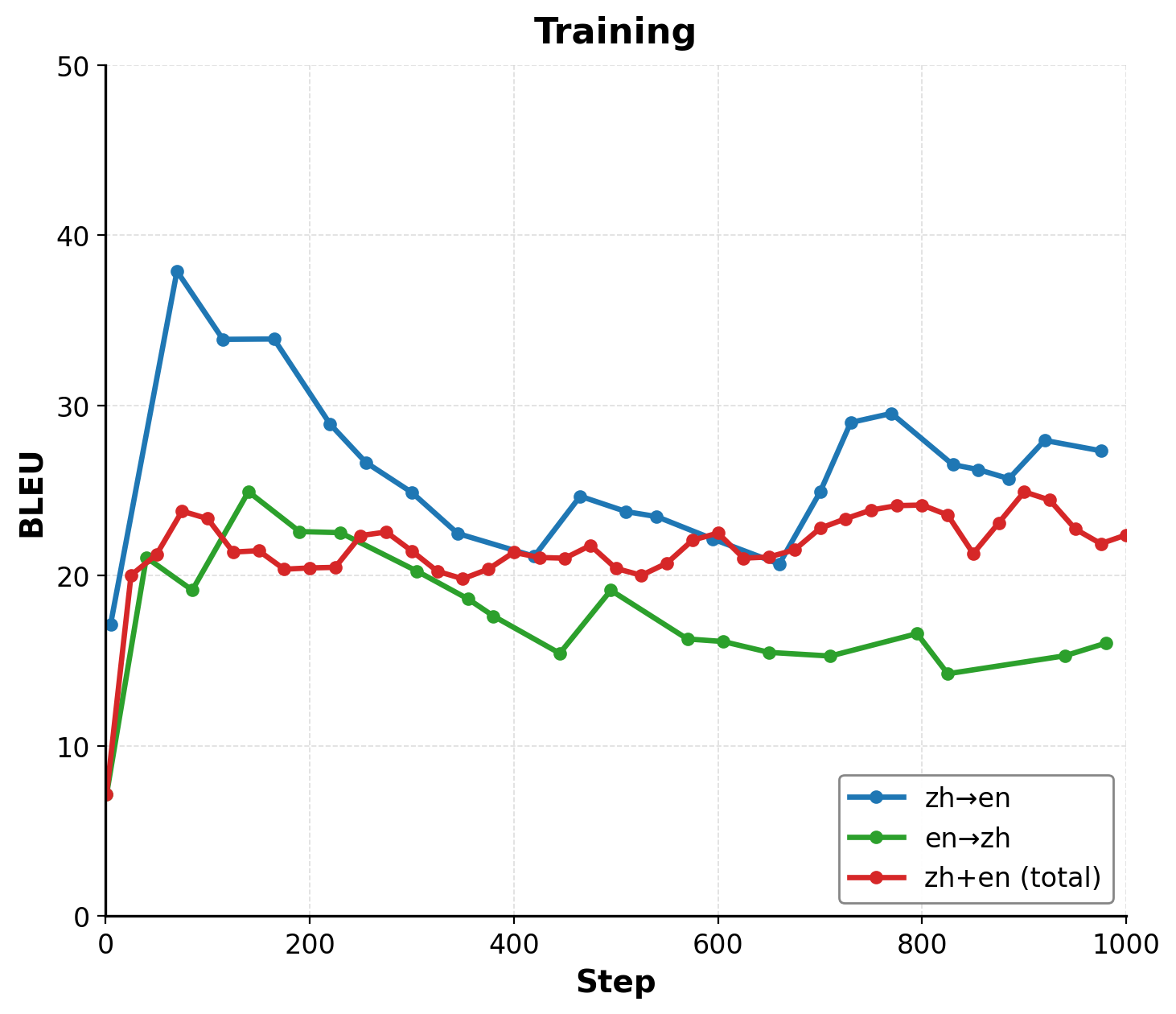}\\
  \small (b) Text-BLEU Reward
\end{minipage}\hfill
\begin{minipage}[t]{0.245\textwidth}
  \centering
  \includegraphics[width=\linewidth]{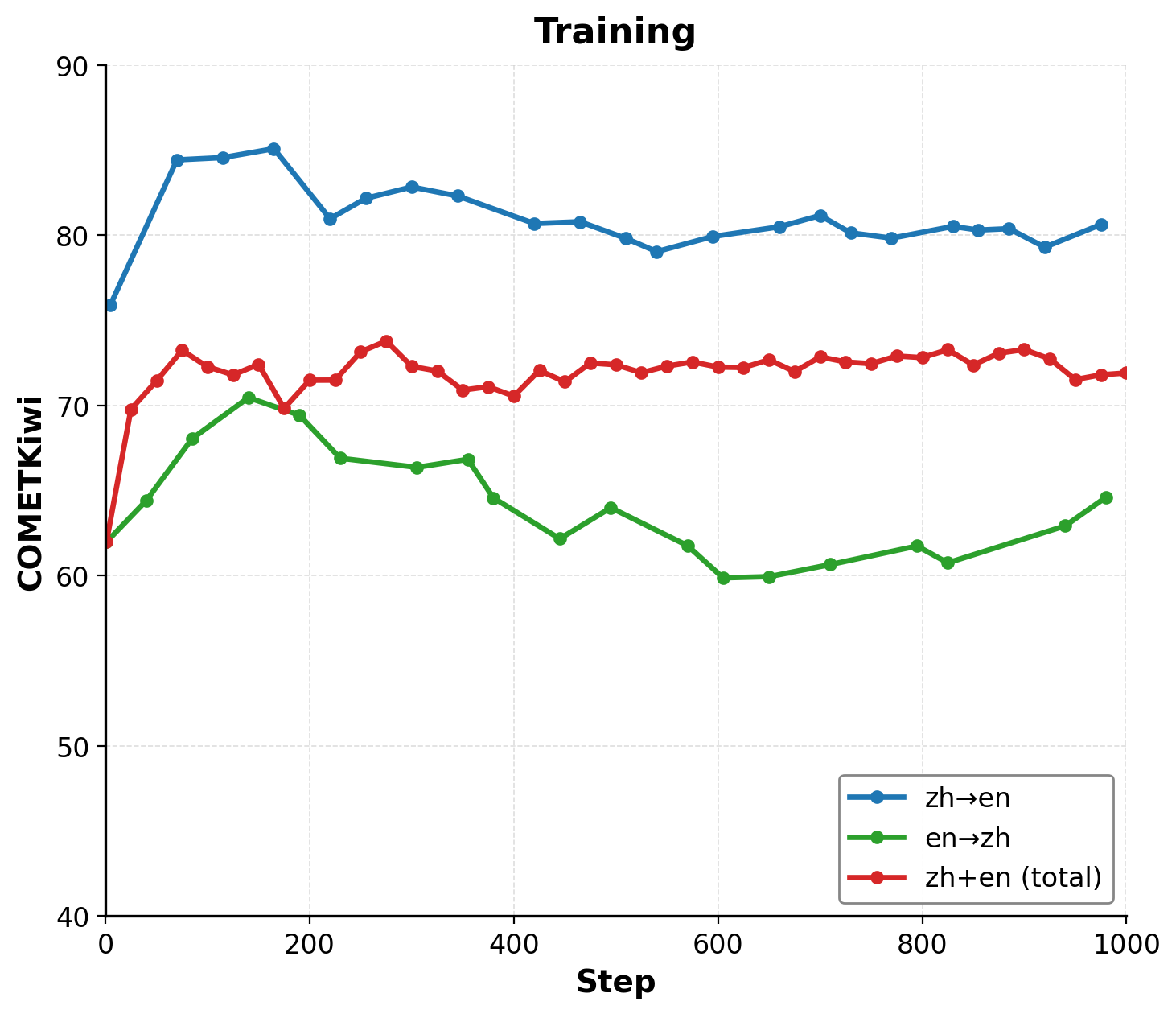}\\
  \small (c) COMET-Kiwi Reward
\end{minipage}\hfill
\begin{minipage}[t]{0.245\textwidth}
  \centering
  \includegraphics[width=\linewidth]{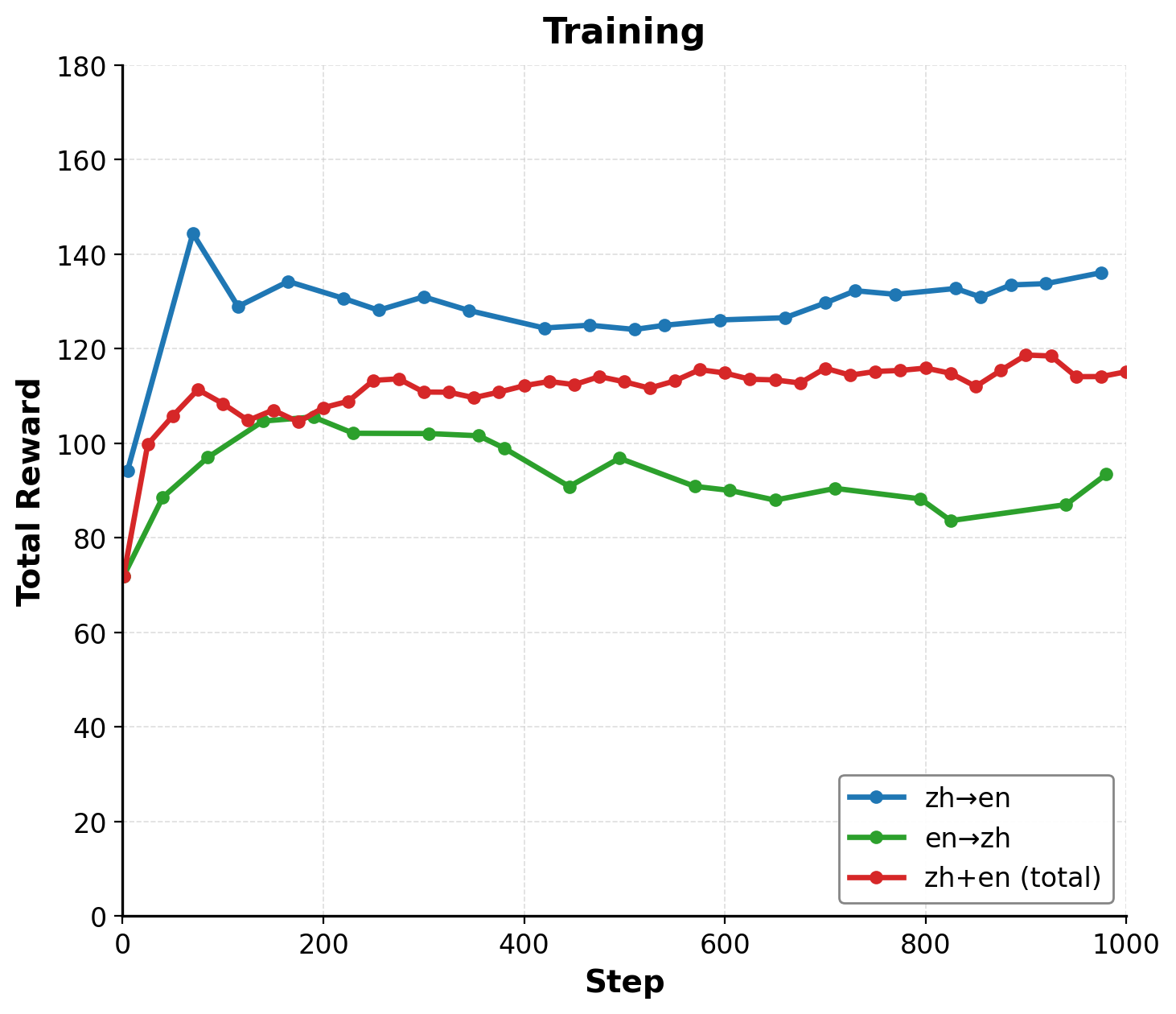}\\
  \small (d) Total Reward
\end{minipage}
\caption{\textbf{Zero-GRPO training dynamics} on CVSS-T. Curves show duration, BLEU, COMET\textsubscript{Kiwi}, and total rewards over $1{,}000$ GRPO steps without DuraSet-440K SFT.}
\label{fig:zero_shot_grpo}
\vspace{-4mm}
\end{figure*}

%% file: tables/mos_results.tex
\begin{table}[t]
\centering

\setlength{\tabcolsep}{0.7mm}
\renewcommand{\arraystretch}{1.15}
\caption{\textbf{Subjective MOS} ($1$--$5$, higher is better) on CVSS-T. Each cell reports \textit{EN-ZH}\,$\vert$\,\textit{ZH-EN}. \textbf{Best}\,/\,\underline{2nd--3rd} per direction are highlighted.}
\label{tab:mos}
\small
\resizebox{\columnwidth}{!}{%
\begin{tabular}{l cccc}
\toprule
\textbf{Model}
 & \textbf{Trans.}$\uparrow$ & \textbf{Natur.}$\uparrow$
 & \textbf{SpkSim}$\uparrow$ & \textbf{Avg}$\uparrow$ \\
\midrule

SeamlessM4T-v2-Large
 & 2.87\,$\vert$\,3.07
 & 1.20\,$\vert$\,3.93
 & 1.13\,$\vert$\,1.27
 & 1.73\,$\vert$\,2.76 \\
Qwen2.5-Omni
 & 3.87\,$\vert$\,3.80
 & \underline{4.17}\,$\vert$\,4.03
 & 1.00\,$\vert$\,2.40
 & 3.01\,$\vert$\,3.41 \\
Kimi-Audio
 & 2.20\,$\vert$\,2.40
 & 3.80\,$\vert$\,3.32
 & 1.07\,$\vert$\,1.13
 & 2.36\,$\vert$\,2.29 \\
UniSS
 & 3.80\,$\vert$\,3.67
 & 3.87\,$\vert$\,3.33
 & 3.87\,$\vert$\,3.93
 & 3.85\,$\vert$\,3.64 \\
\midrule
\rowcolor{gray!15}
\textbf{DuraS2ST-Baseline}
 & 3.93\,$\vert$\,3.53
 & 4.10\,$\vert$\,\underline{4.13}
 & 4.00\,$\vert$\,\textbf{4.33}
 & 4.01\,$\vert$\,4.00 \\
\rowcolor{gray!15}
\textbf{DuraS2ST-Instruct}
 & \underline{4.00}\,$\vert$\,\underline{4.06}
 & \underline{4.20}\,$\vert$\,\underline{4.08}
 & \underline{4.13}\,$\vert$\,\underline{4.00}
 & \underline{4.11}\,$\vert$\,\underline{4.05} \\
\rowcolor{gray!15}
\textbf{DuraS2ST-Think}
 & \textbf{4.53}\,$\vert$\,\underline{4.13}
 & \textbf{4.27}\,$\vert$\,\underline{4.13}
 & \underline{4.27}\,$\vert$\,\underline{4.20}
 & \textbf{4.36}\,$\vert$\,\underline{4.15} \\
\rowcolor{gray!15}
\textbf{DuraS2ST-Think-RL}
 & \underline{4.33}\,$\vert$\,\textbf{4.73}
 & \underline{4.17}\,$\vert$\,\textbf{4.23}
 & \textbf{4.40}\,$\vert$\,\textbf{4.33}
 & \underline{4.30}\,$\vert$\,\textbf{4.43} \\
 
\bottomrule
\end{tabular}}
\vspace{-4mm}
\end{table}

%% file: tables/zero_grpo_duration.tex
\begin{table}[t]
\centering
\setlength{\tabcolsep}{0.9mm}
\renewcommand{\arraystretch}{1.15}
\caption{\textbf{Zero-GRPO duration consistency} on CVSS-T (EN-ZH\,$\vert$\,ZH-EN). Removing $R_{\text{dur}}$ isolates the effect of Duration Margin Reward without DuraSet-440K SFT.}
\label{tab:zero_grpo_duration}
\small
\resizebox{\columnwidth}{!}{%
\begin{tabular}{l cccc}
\toprule
\textbf{Model}
 & \textbf{SLC-0.2}$\uparrow$ & \textbf{SLC-0.4}$\uparrow$
 & \textbf{MADE\,(s)}$\downarrow$ & \textbf{MRDE}$\downarrow$ \\
\midrule
Baseline
 & 0.487\,$\vert$\,0.305 & 0.851\,$\vert$\,0.680
 & 2.04\,$\vert$\,2.22 & 0.629\,$\vert$\,0.396 \\
\rowcolor{gray!15}
\textbf{+\,GRPO}
 & 0.455\,$\vert$\,0.302 & \textbf{0.877}\,$\vert$\,\textbf{0.691}
 & \textbf{1.15}\,$\vert$\,\textbf{1.97} & \textbf{0.259}\,$\vert$\,\textbf{0.346} \\
 \quad w/o $R_{{dur}}$
 & 0.448\,$\vert$\,0.319 & 0.855\,$\vert$\,0.641
 & 1.83\,$\vert$\,1.93 & 0.565\,$\vert$\,0.385 \\
\bottomrule
\end{tabular}}
\vspace{-4mm}
\end{table}

%% file: sections/conclusion.tex
\vspace{-2mm}
\section{Conclusion}
\label{sec:conclusion}
\vspace{-2mm}
We presented DuraS2ST, a reasoning-based framework for duration-aligned speech-to-speech translation. Instead of treating duration control as acoustic post-processing, DuraS2ST explicitly plans target wording and phonetic length before generating the final interleaved text-acoustic sequence. To support this paradigm, we constructed DuraSet-440K, a large-scale duration-aligned reasoning corpus, and trained the model in two phases of supervised fine-tuning followed by multi-modal GRPO. Within the RL phase, the Duration Margin Reward resolves the quality-duration trade-off through a saturated soft constraint, and the Modality-Aware Reward Attribution prevents cross-modality reward contamination by restricting the duration advantage to acoustic tokens. Experiments on CVSS-T show that DuraS2ST balances translation quality and duration consistency. A zero-SFT ablation further shows that our rewards substantially reduce MADE, highlighting their complementary contribution beyond duration-aligned SFT.

%% file: sections/limitations.tex
\section*{Limitations}
\label{sec:conclusion}
We proposed DuraS2ST, a chain-of-thought and reinforcement learning framework for duration-aligned speech-to-speech translation, and constructed DuraSet-440K to support its training. However, duration alignment in S2ST is more nuanced than a single language pair can fully reveal, so our current evaluation has focused on bidirectional Chinese-English translation, a representative pair of structurally divergent languages, laying the groundwork for broader linguistic validation. In addition, the current think-then-speak architecture decodes the chain-of-thought rationale in full before emitting any acoustic token, which fits offline applications such as video dubbing but does not yet support streaming or simultaneous interpretation. Future work will both extend DuraS2ST and DuraSet-440K to additional language pairs and explore streaming variants in which chain-of-thought reasoning and acoustic emission interleave at finer granularity.

\section*{Ethical Considerations}

This work studies duration-aligned speech-to-speech translation with synthetic speech generation and speaker preservation. While such techniques can improve accessibility and cross-lingual communication, they may also be misused for impersonation, unauthorized voice conversion, or misleading audio generation. We therefore emphasize that the proposed system should only be applied to speech data with proper consent, licensing, and appropriate safeguards. The datasets used in our experiments are public research datasets, and our evaluations are conducted on anonymized samples. We use public research datasets and existing models in accordance with their licenses and terms of use. Human evaluation participants are asked to assess translation adequacy, naturalness, and speaker similarity, without being exposed to personally identifying information beyond the speech content available in the benchmark data. We also acknowledge that our experiments focus on English and Chinese, and the findings may not fully generalize to other languages, accents, or sociolinguistic settings. 

%% file: sections/acknowledgement.tex
\section*{Acknowledgments}
\label{sec:acknowledgments}
This work is partially supported by the General Research Fund from the Research Grants Council of Hong Kong SAR Government (Project No. 14202623).

%% file: sup-sections/additional_experiments.tex
\section{Additional Experiments}
\label{sec:appendix-additional}

This section reports auxiliary experiments that support the training design choices of DuraS2ST.

\subsection{Pre-Rollout Filtering}
\label{sec:appendix-prerollout}

GRPO relies on within-group reward variation to estimate group-relative advantages. If all rollouts for a prompt receive nearly identical rewards, the prompt provides little useful gradient signal but still consumes rollout budget. To improve training efficiency, we perform a one-time \emph{pre-rollout filtering} step before GRPO.

\paragraph{Setup.}
For each candidate prompt, we sample $G=8$ rollouts from the SFT checkpoint using the same decoding configuration as GRPO. We discard prompts with near-zero reward variance, since they cannot provide informative group-relative advantages. The remaining prompts are bucketed into \emph{easy}, \emph{medium}, and \emph{hard} groups according to their mean total reward. The final GRPO training set uses all medium prompts and a small fraction of easy prompts, balancing learnability and stability.

\paragraph{Result.}
We compare two GRPO runs with identical rewards, rollout budget, and hyperparameters, differing only in the training prompts: the original candidate pool and the pre-rollout-filtered set. Figure~\ref{fig:ablation_prerollout} shows the reward curves on ZH$\to$EN. Pre-rollout filtering yields faster improvement and a higher reward plateau across duration, translation, and total rewards. This indicates that filtering out uninformative prompts makes GRPO more sample-efficient and stabilizes optimization.

\input{figures/ablation_prerollout}

%% file: figures/ablation_prerollout.tex
\begin{figure*}[t]
\centering
\begin{minipage}[t]{0.245\textwidth}
  \centering
  \includegraphics[width=\linewidth]{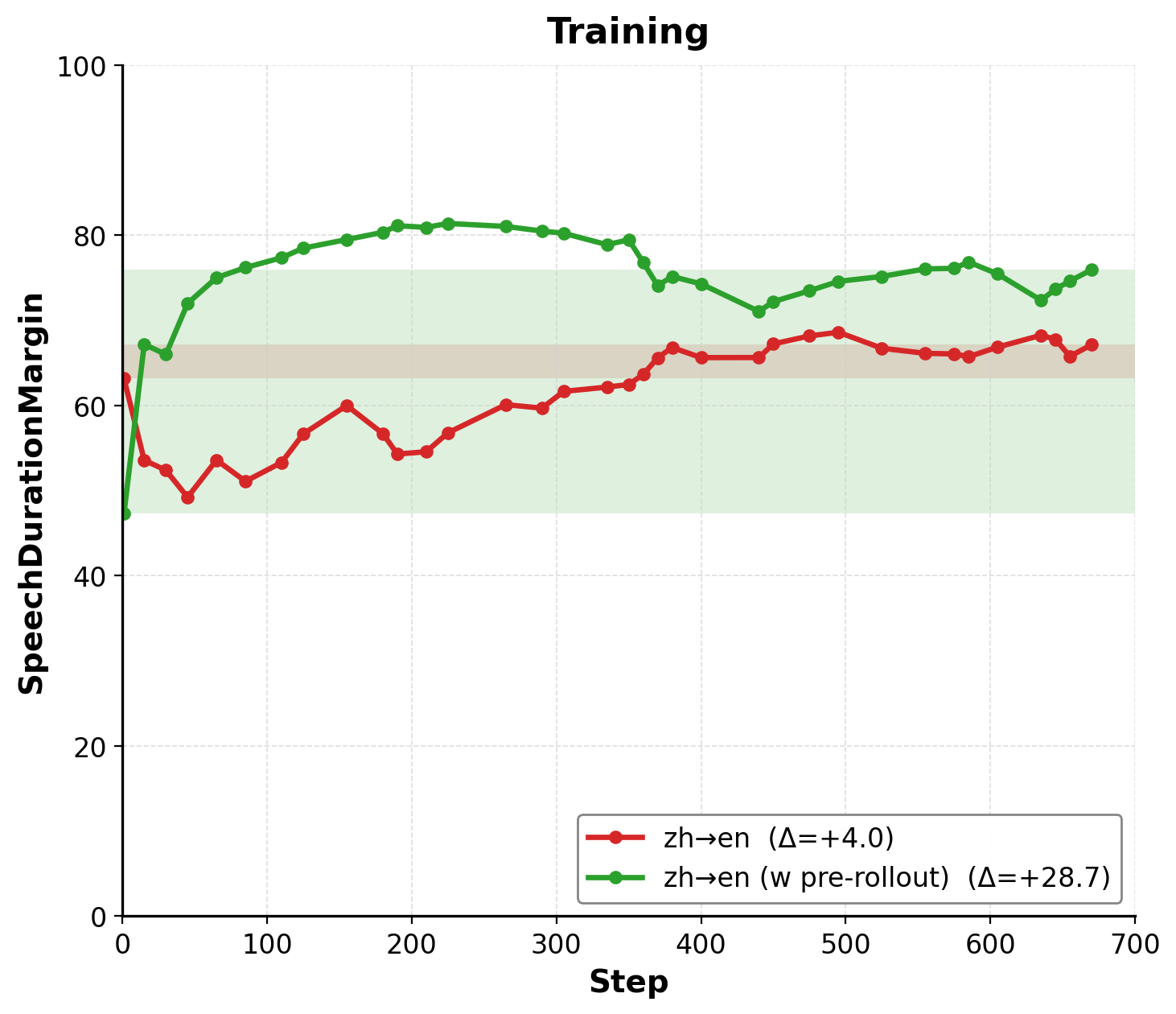}\\
  \small (a) Duration Margin Reward
\end{minipage}\hfill
\begin{minipage}[t]{0.245\textwidth}
  \centering
  \includegraphics[width=\linewidth]{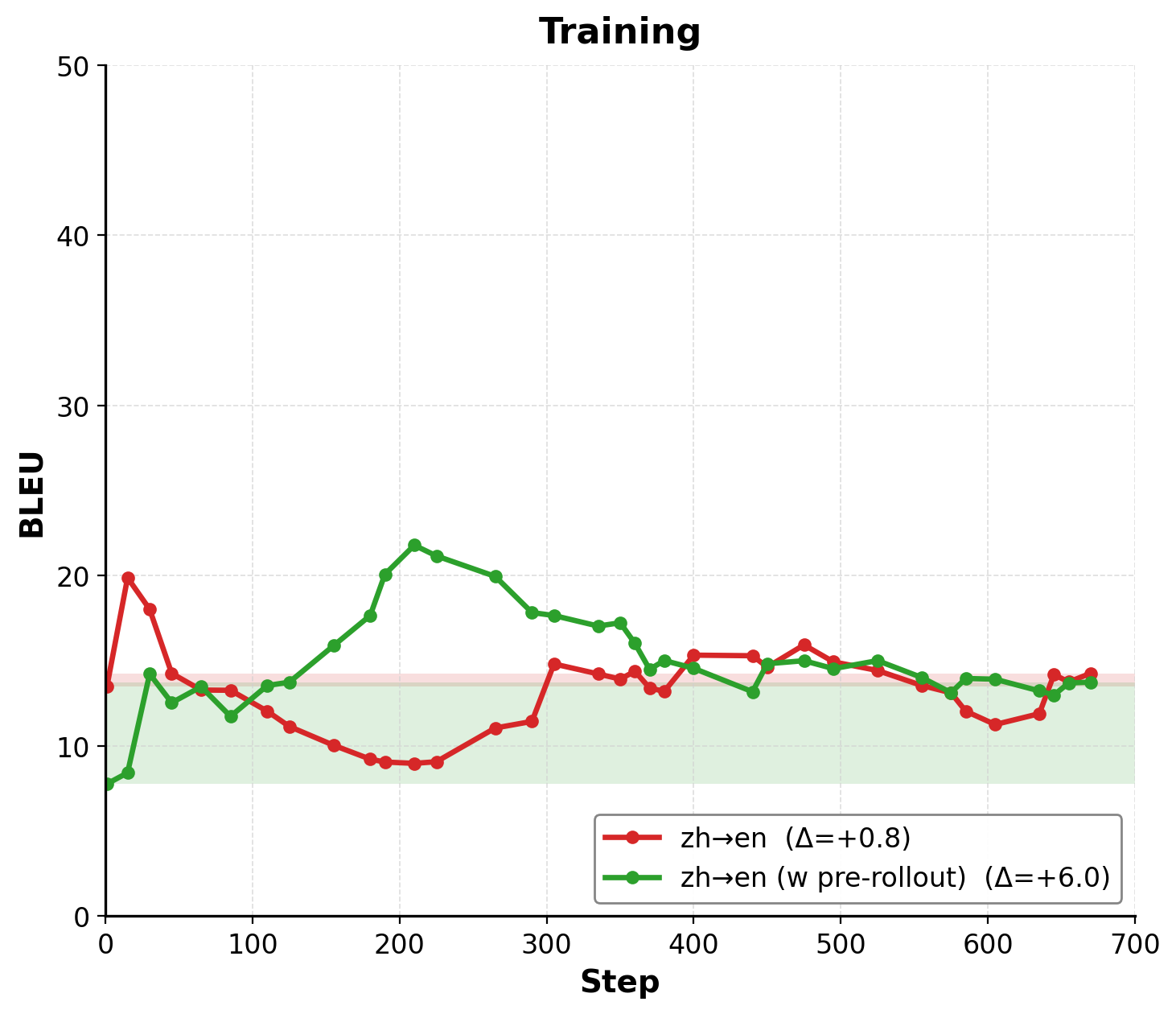}\\
  \small (b) Text-BLEU Reward
\end{minipage}\hfill
\begin{minipage}[t]{0.245\textwidth}
  \centering
  \includegraphics[width=\linewidth]{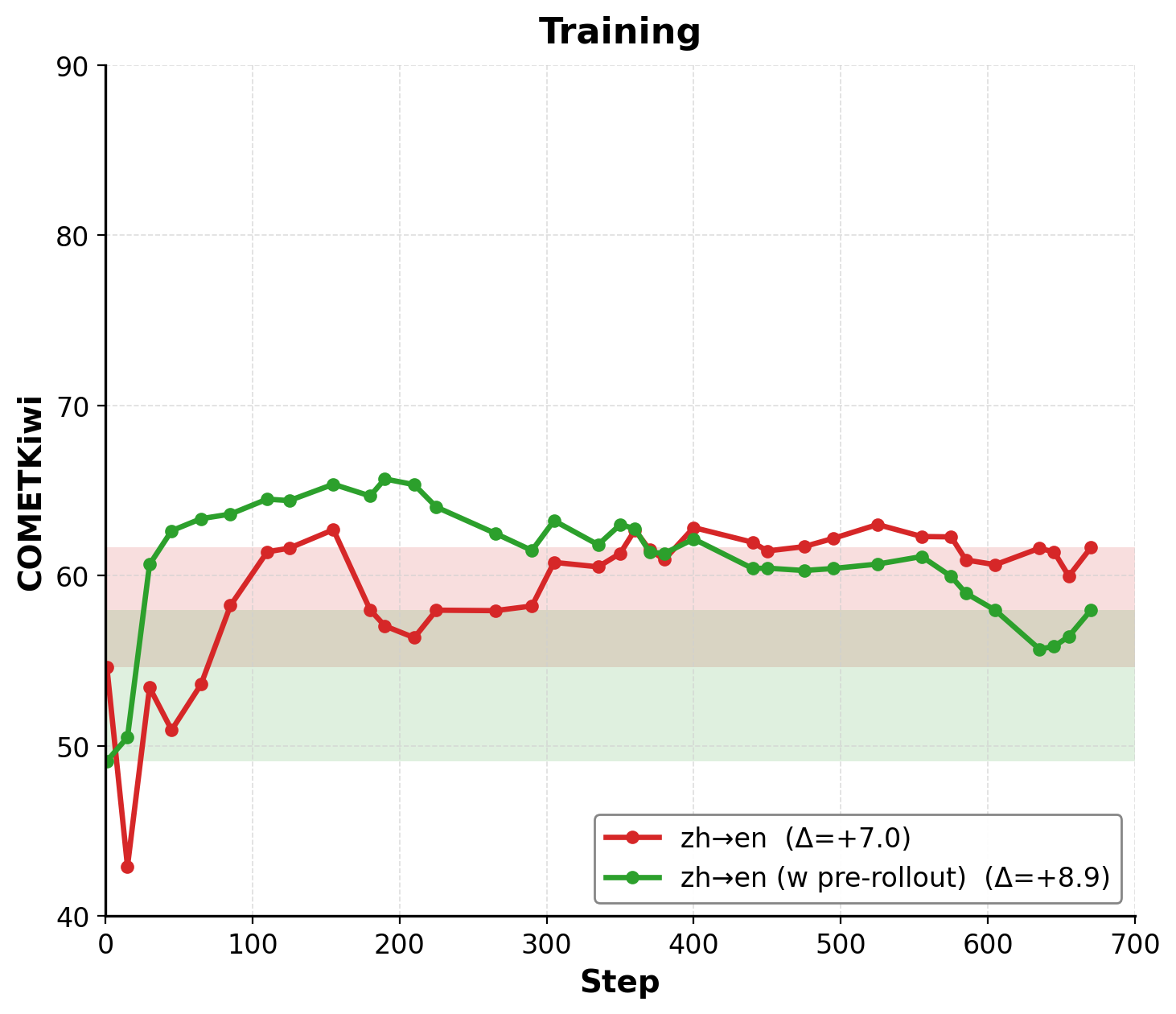}\\
  \small (c) COMET-Kiwi Reward
\end{minipage}\hfill
\begin{minipage}[t]{0.245\textwidth}
  \centering
  \includegraphics[width=\linewidth]{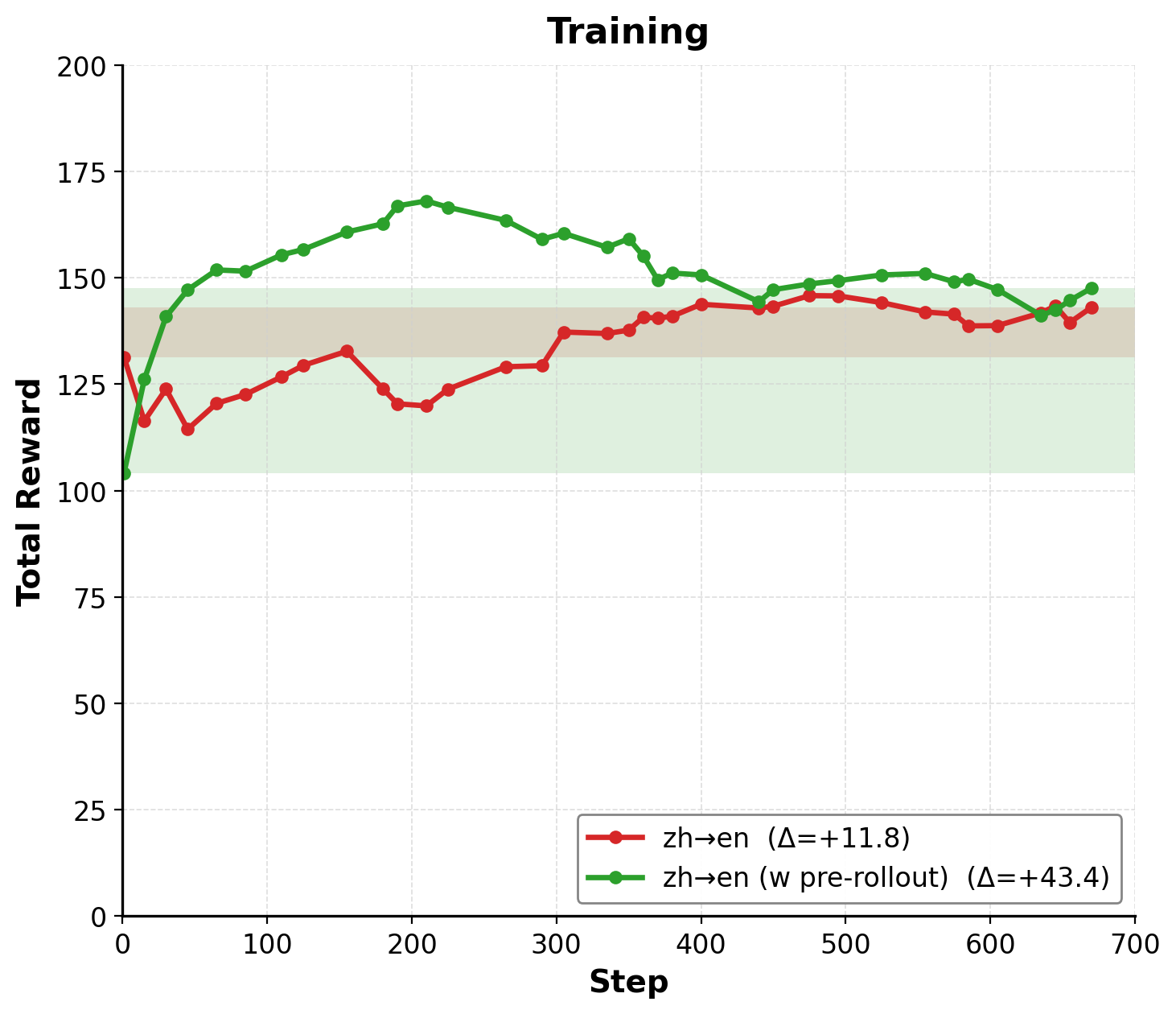}\\
  \small (d) Total Reward
\end{minipage}
\caption{\textbf{Pre-rollout difficulty filtering improves GRPO stability and efficiency.} On ZH$\to$EN, two GRPO runs share an identical reward composition, rollout budget, and hyperparameters, differing only in their training data. Across all four reward components, the pre-rollout-filtered run (green) climbs faster and reaches a higher plateau than the unfiltered baseline (red); the in-legend $\Delta$ reports the improvement from the start of training to step~700.}
\label{fig:ablation_prerollout}
\end{figure*}

%% file: sup-sections/duraset.tex
\section{DuraSet-440K Details}
\label{sec:appendix-dataset}

\begin{figure*}[ht]
    \centering
    \includegraphics[width=\linewidth]{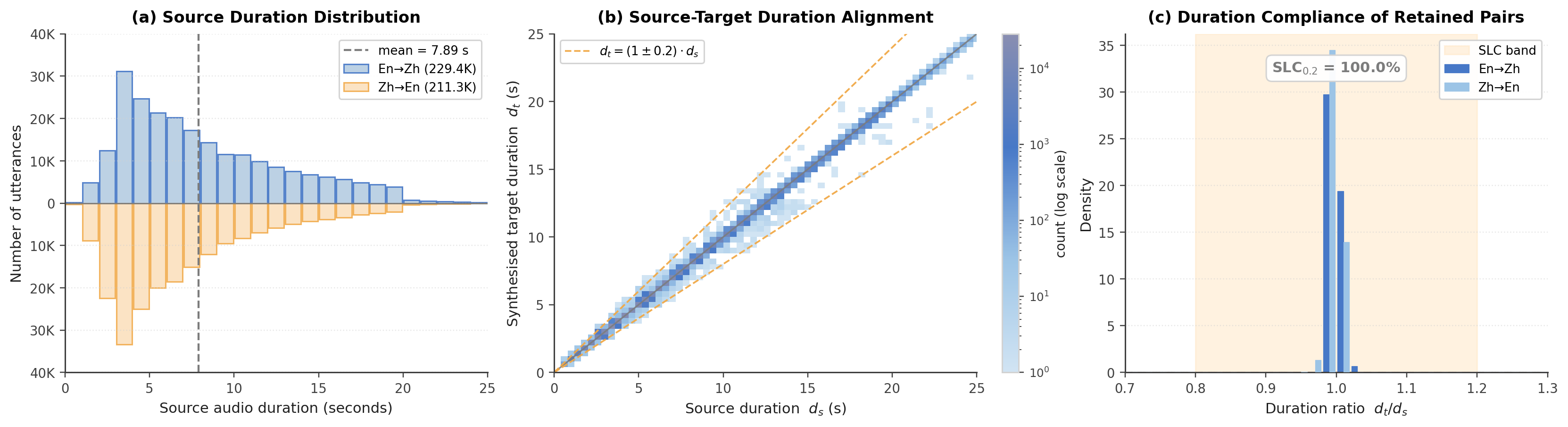}
    \caption{\textbf{Duration statistics of DuraSet-440K.} (a) Source-duration histogram by direction. (b) Hex-binned joint density of source and target durations with the $d_{\mathrm{tgt}}=d_{\mathrm{src}}$ diagonal. (c) Distribution of the target-to-source duration ratio $d_{\mathrm{tgt}}/d_{\mathrm{src}}$.}
    \label{fig:duration-dist}
\end{figure*}
This section provides the technical details of the DuraSet-440K construction pipeline, including model choices, hyperparameters, filtering criteria, and prompting templates. It complements the overview in Section~\ref{sec:dataset}.

\subsection{Translation Candidate Construction}
\label{sec:appendix-translation}

For each source utterance, we first construct a set of translation candidates with diverse lexical densities and phonetic lengths. This candidate set provides a search space for selecting translations that preserve the source meaning while better matching the target duration budget.

\begin{figure*}[ht]
    \centering
    \includegraphics[width=0.85\linewidth]{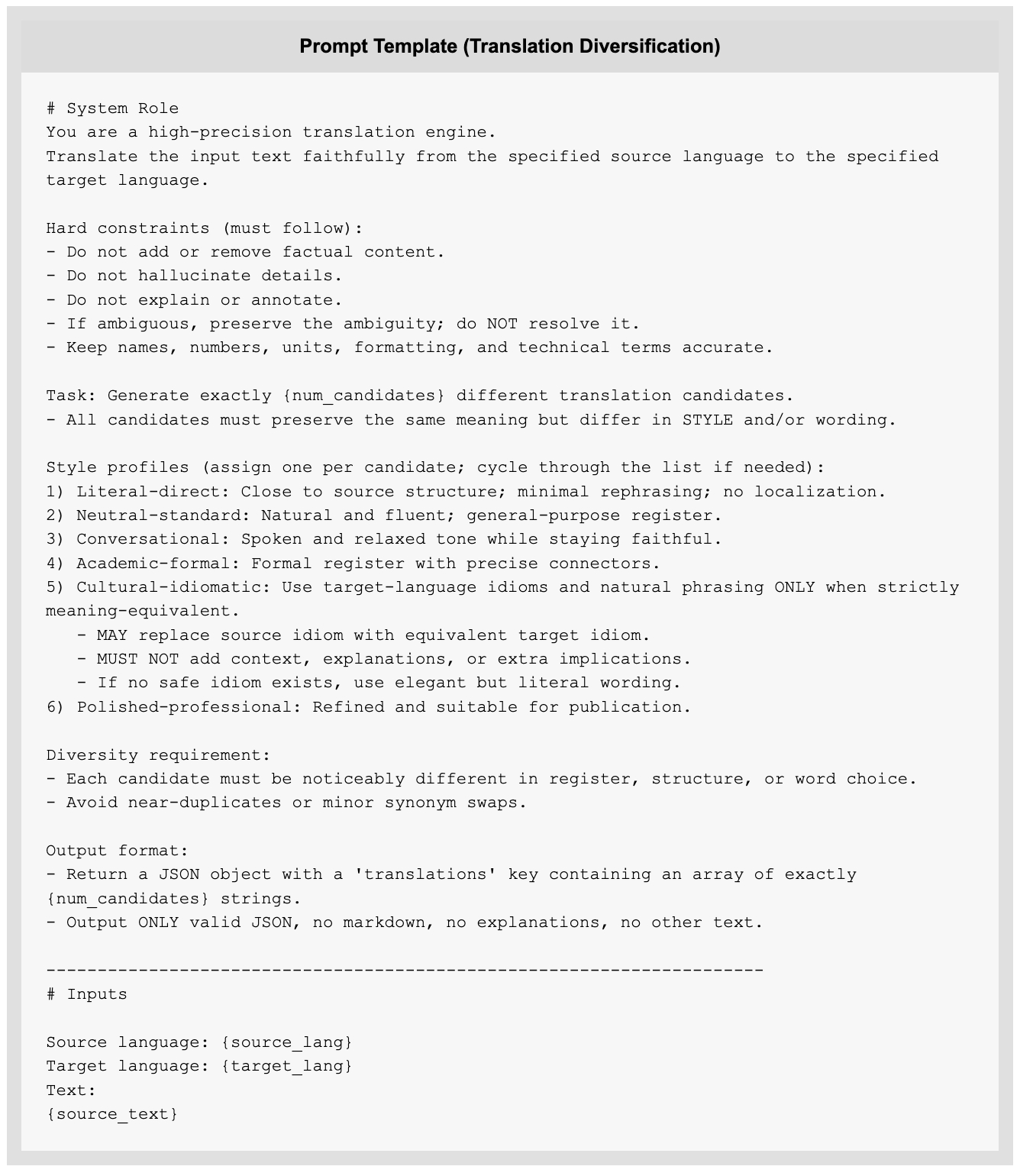} 
    \caption{Prompt template for translation candidate construction. The prompt encourages stylistically diverse translations with different lexical densities and temporal footprints for the same source utterance.}
    \label{fig:prompt_translation}
\end{figure*}

We use GPT-4o to generate translation candidates with temperature $T=0.7$ and $\mathrm{top\text{-}}p=0.9$. For each source utterance, we sample $N=5$ candidates using a structured prompt that encourages stylistic diversity. As shown in Figure~\ref{fig:prompt_translation}, the prompt assigns distinct translation styles ranging from verbose to concise, encouraging the LLM to vary wording, syllable count, and phonetic footprint while preserving the core semantics.

\subsection{Target Speech Synthesis Configuration}
\label{sec:appendix-synthesis}


We synthesize each translation candidate using an in-house TTS model with zero-shot duration control at 24\,kHz. The source audio serves as the timbre prompt, allowing the synthesized target speech to preserve the speaker characteristics of the source utterance.

For each translation candidate $\mathbf{T}_i$, we estimate its natural duration $d_{\mathrm{nat}}^{(i)}$ and compute a global frame scaling factor against the target duration budget $d_{\mathrm{budget}}$:
\begin{equation}
s_i = \mathrm{clip}\!\left(\frac{d_{\mathrm{budget}}}{d_{\mathrm{nat}}^{(i)}}, \; 0.5, \; 2.0\right).
\end{equation}
The predicted acoustic frames are scaled by $s_i$ before waveform generation. The clipping range $[0.5,2.0]$ prevents excessive compression or expansion, which could otherwise degrade speech quality. It also ensures that candidates with incompatible phonetic lengths retain measurable residual duration errors, allowing the filtering stage to reject them.

During synthesis, we use a classifier-free guidance (CFG) scale no larger than $10$ for stable generation. The source timbre prompt is trimmed and padded with a $0.2$\,s leading fade and a $0.5$\,s trailing fade to smooth conditioning transitions.

\subsection{Multi-Dimensional Filtering}
\label{sec:appendix-filtering}

Each synthesized candidate is filtered by transcription fidelity, duration alignment, and speaker similarity. A candidate $\mathbf{Y}_i$ is retained only if it satisfies:
\begin{equation}
\begin{aligned}
\mathrm{WER}(\mathbf{Y}_i, \mathbf{T}_i) &\le \tau_{\mathrm{wer}}, \\
|\Delta d_i| &\le \tau_{\mathrm{abs}}, \\
\frac{|\Delta d_i|}{d_{\mathrm{src}}} &\le \tau_{\mathrm{rel}}, \\
\cos(\mathbf{e}_{\mathrm{src}}, \mathbf{e}_{\mathrm{tgt}}^{(i)}) &\ge \tau_{\mathrm{sim}},
\end{aligned}
\label{eq:quality_filter}
\end{equation}
where $\Delta d_i = d_{\mathrm{tgt}}^{(i)} - d_{\mathrm{src}}$, $d_{\mathrm{tgt}}^{(i)}$ is the duration of the synthesized candidate, and $\mathbf{e}_{\mathrm{src}}$ and $\mathbf{e}_{\mathrm{tgt}}^{(i)}$ are speaker embeddings of the source and target speech.

We set the thresholds to $\tau_{\mathrm{wer}}=0.05$, $\tau_{\mathrm{abs}}=5\,\mathrm{s}$, $\tau_{\mathrm{rel}}=0.20$, and $\tau_{\mathrm{sim}}=0.86$. WER is computed using Fun-ASR-Nano-2512 with inverse text normalization (ITN) enabled, which standardizes numbers and punctuation before text alignment. Speaker similarity is computed with WavLM-base-plus-sv embeddings extracted at 16\,kHz.

Among candidates that pass all filtering criteria, we select the final target speech by minimizing duration error, using speaker similarity and WER as secondary tie-breakers. This selection procedure favors duration-aligned pairs while maintaining transcription accuracy and speaker consistency.

\subsection{Reasoning Rationale Construction}
\label{sec:appendix-reasoning}

After selecting the duration-aligned target pair, we annotate each example with a phoneme-grounded reasoning rationale. The goal is to provide explicit supervision for forward planning: before generating target speech, the model should reason about translation choices, phonetic length, and the duration budget.

\begin{figure*}[ht]
    \centering
    \includegraphics[width=0.85\linewidth]{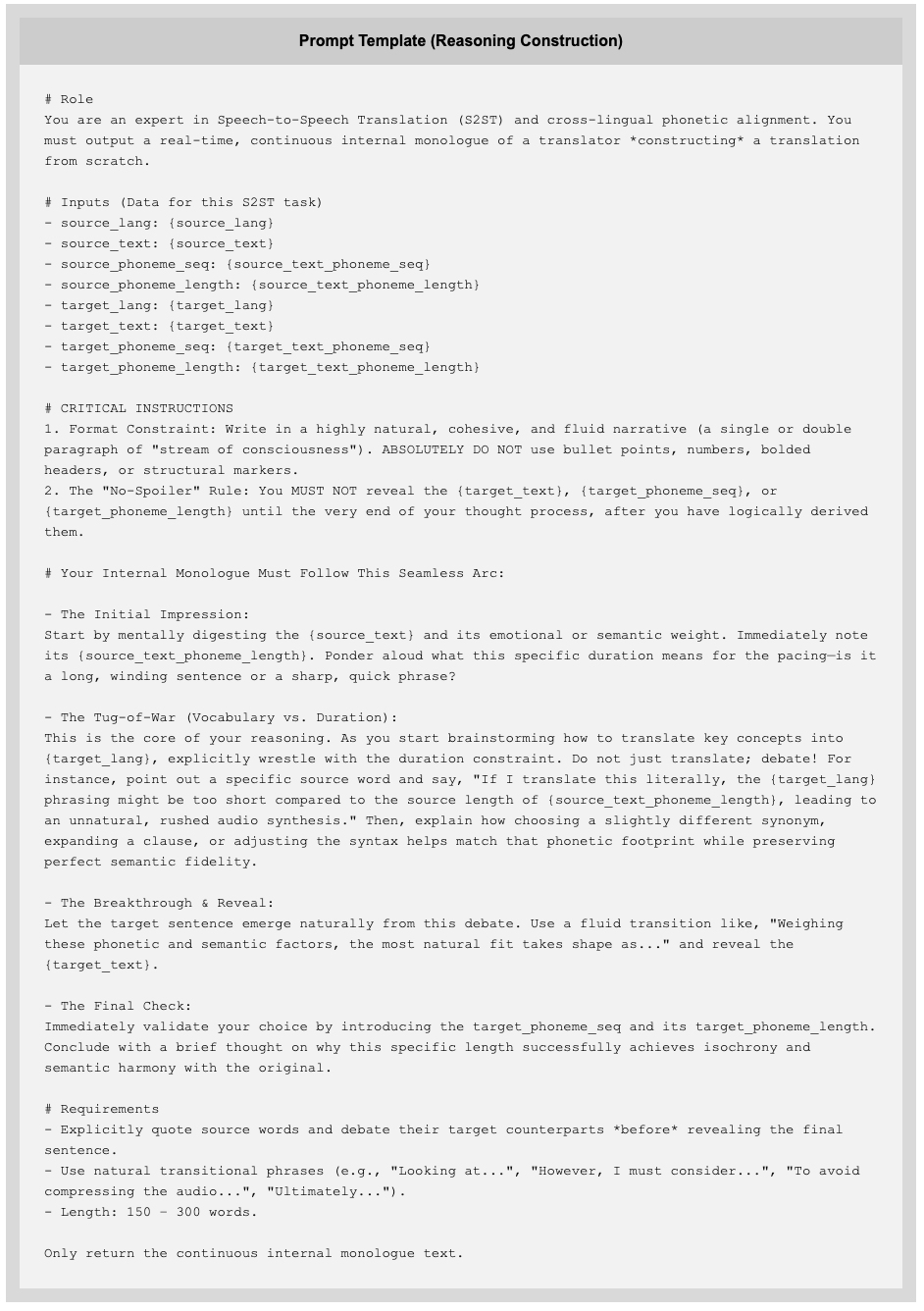}
    \caption{Prompt template for reasoning rationale construction. The prompt guides the LLM to compare translation choices against deterministic phoneme lengths before outputting the final translation.}
    \label{fig:prompt_reasoning}
\end{figure*}

We query GPT-4o with temperature $T=0.0$ to obtain deterministic rationales. Each rationale is constrained to 150--300 words to maintain sufficient detail while avoiding unnecessary verbosity. As shown in Figure~\ref{fig:prompt_reasoning}, the prompt follows a structured reasoning template and enforces a ``no-spoiler'' rule, requiring the model to discuss translation alternatives before revealing the final target text.

To ground the rationale in verifiable phonetic evidence, we provide the source and target phoneme sequences together with their lengths. These phoneme sequences are deterministically extracted using the BigCiDian lexicon with a longest-match strategy and mapped to the International Phonetic Alphabet (IPA). This forces the rationale to explicitly connect wording choices with phonetic footprints and duration constraints.

\subsection{Dataset Statistics}
\label{sec:appendix-stats}

The final DuraSet-440K corpus contains 440{,}705 parallel utterances, including 229{,}410 En$\to$Zh examples and 211{,}295 Zh$\to$En examples. Figure~\ref{fig:duration-dist} summarizes the duration statistics of the retained corpus.

Figure~\ref{fig:duration-dist}(a) shows the distribution of source durations, with a mean of $7.89$\,s and a standard deviation of $4.68$\,s. Figure~\ref{fig:duration-dist}(b) plots the joint density of source and target durations, where most examples concentrate around the $d_{\mathrm{tgt}}=d_{\mathrm{src}}$ diagonal. Figure~\ref{fig:duration-dist}(c) shows the distribution of target-to-source duration ratios, which is tightly centered around $\mu=0.995$ with $\sigma=0.008$ and achieves $\mathrm{SLC}_{0.2}=100.0\%$. These statistics confirm that DuraSet-440K provides temporally aligned source-target pairs suitable for training duration-aware S2ST models.

%% file: sup-sections/implementation_details.tex
\section{Implementation Details}
\label{app:implementation}

This section provides additional implementation details for DuraS2ST, including training hyperparameters, reward computation, baseline inference protocols, and evaluation metrics.

\subsection{Training Hyperparameters}
\label{app:training}

\paragraph{Phase I: SFT.}
We perform full-parameter fine-tuning of Step-Audio-2-mini-Think on $8$ NVIDIA A100 80\,GB GPUs with bf16 precision and DeepSpeed ZeRO-3. We use AdamW with a learning rate of $1\mathrm{e}{-5}$, $5\%$ linear warmup, a per-device batch size of $1$, and gradient accumulation of $16$, resulting in a global batch size of $128$. The maximum sequence length is set to $16{,}384$ tokens to accommodate the interleaved CoT and acoustic-token streams. The model is trained for $4{,}000$ steps using the CoT-then-speech format defined in Eq.~\ref{eq:output_format}.

\paragraph{Phase II: GRPO.}
Starting from the SFT checkpoint, we attach LoRA adapters with rank $16$ and $\alpha=32$ to all linear projections, and optimize the model with GRPO for $1{,}000$ steps. We use a learning rate of $1.5\mathrm{e}{-5}$ and a KL coefficient of $\beta=0.001$ against the SFT reference policy. For each input, we sample $G=16$ rollouts with temperature $1.0$, top-$p=0.99$, and top-$k=50$. Rollout generation is served by vLLM~\citep{kwon2023efficient} to reduce the cost of long autoregressive speech decoding. Following DAPO~\citep{yu2026dapo}, groups whose rewards collapse to a single value are re-sampled up to $3$ times, and rollouts longer than $4{,}096$ tokens are filtered before the gradient update.

\subsection{RL Reward Implementation}
\label{app:reward}

All rewards are computed online from the generated token stream, without invoking ASR during training. Step-Audio-2-mini-Think emits an interleaved \emph{TA4} stream, where one text token is placed before every four acoustic tokens. The text tokens inside the \texttt{<tts>} segment therefore form a temporally aligned transcript of the generated speech. We concatenate these in-segment text tokens as the post-CoT translation $\mathbf{y}_{\text{text}}$ for reward computation.

\paragraph{BLEU reward.}
We compute sentence-level SacreBLEU~\citep{post2018sacrebleu} between $\mathbf{y}_{\text{text}}$ and the reference translation $\mathbf{y}^*$. The score is divided by $100$ to obtain a reward in $[0,1]$. Language-specific preprocessing follows the evaluation pipeline in Appendix~\ref{app:metric}.

\paragraph{COMET\textsubscript{Kiwi} reward.}
We compute the reference-free semantic reward using \texttt{Unbabel/wmt22-cometkiwi-da}~\citep{rei2022cometkiwi}, with the source text and generated translation as input. The model score is used directly as the COMET\textsubscript{Kiwi} reward.

\paragraph{Duration Margin Reward.}
The duration reward is computed from the acoustic-token length. Given the predicted length $L_{\text{pred}}$ and target length $L_{\text{tgt}}$, we compute the relative duration gap
$\delta = |L_{\text{pred}} - L_{\text{tgt}}| / L_{\text{tgt}}$
and apply the Duration Margin Reward. We use tolerance $\tau=0.05$, smoothing scale $s=0.02$, scaling factor $\kappa=0.5$, and margin clipping bound $B=5$. Under this setting, the reward is high for sub-$5\%$ duration gaps, concentrates its learning signal around the recoverable $5\%$--$15\%$ range, and saturates for larger deviations. These constants are tuned on a small held-out subset of DuraSet-440K.

\paragraph{Format gate.}
Each rollout is checked by a deterministic parser. A rollout is valid only if the control tags \texttt{<think>}, \texttt{</think>}, and \texttt{<tts\_start>} appear in the required order, with no trailing tokens after \texttt{<tts\_end>}. Valid rollouts receive $g=1$, while malformed rollouts receive $g=0$.

\paragraph{Gated aggregation and attribution.}
As described in Section~\ref{subsubsec:reward_design}, the format gate multiplicatively zeroes both the quality and duration rewards before group-relative normalization. This prevents malformed outputs from receiving positive credit for partially satisfying translation or duration objectives. The gated rewards are then standardized into per-reward advantages and assigned to token spans through MARA: the quality advantage is applied to both reasoning and acoustic spans, while the duration advantage is restricted to acoustic-phase tokens.

\subsection{Baselines}
\label{app:baseline}

For each external baseline, we use the model's default decoding configuration and a fixed direction-specific translation instruction. We do not apply additional input preprocessing or output post-processing, so the reported scores reflect each system's raw end-to-end S2ST behavior under a controlled prompt.

\paragraph{SeamlessM4T.}
We evaluate SeamlessM4T~\citep{barrault2023seamless} using three publicly released checkpoints: \texttt{facebook/hf-seamless-m4t-medium}, \texttt{facebook/hf-seamless-m4t-large}, and \texttt{facebook/seamless-m4t-v2-large}. Inference is performed with the official Hugging Face Transformers wrappers and default decoding parameters. The target-language token is set to \texttt{eng} for ZH$\to$EN and \texttt{cmn} for EN$\to$ZH.

\paragraph{Step-Audio-2-mini and Step-Audio-2-mini-Think.}
The official Step-Audio-2 release provides the following Chinese-target S2ST prompt:
\begin{tcolorbox}[colback=gray!5, colframe=black!40, boxrule=0.4pt, arc=2pt]
\footnotesize
\zh{请仔细聆听这段语音，然后将其内容翻译成中文并用语音播报。}
\end{tcolorbox}
For ZH$\to$EN, we find that simply replacing the target language with English may lead to a mismatch between the reasoning text and the generated speech. We therefore explicitly specify both the source and target languages:
\begin{tcolorbox}[colback=gray!5, colframe=black!40, boxrule=0.4pt, arc=2pt]
\footnotesize
\zh{请仔细聆听这段中文语音，然后将其内容翻译成英文并用英文语音播报。}
\end{tcolorbox}
We use the original Chinese-target prompt for EN$\to$ZH and the dual-language English-target prompt for ZH$\to$EN. The same prompt templates are used for DuraSet-440K construction and DuraS2ST evaluation.

\paragraph{Qwen2.5-Omni.}
We evaluate Qwen2.5-Omni~\citep{xu2025qwen25omni} in its default S2ST mode with a minimal translation instruction following~\citet{cheng2025uniss}. For example:
\begin{tcolorbox}[colback=gray!5, colframe=black!40, boxrule=0.4pt, arc=2pt]
\footnotesize
\texttt{Translate the following Chinese speech into English.}
\end{tcolorbox}
and the symmetric instruction for EN$\to$ZH. We observe that Qwen2.5-Omni sometimes appends conversational fillers after the translated speech, which are retained in the generated waveform and affect both ASR-based translation metrics and duration metrics. We do not remove these artifacts by post-processing.

\paragraph{Kimi-Audio.}
We use the official \texttt{Kimi-Audio-7B-Instruct} checkpoint~\citep{ding2025kimi} with its bundled inference pipeline and recommended decoding configuration. We use the following direction-specific instructions:
\begin{tcolorbox}[colback=gray!5, colframe=black!40, boxrule=0.4pt, arc=2pt]
\footnotesize
\texttt{Translate the above Chinese audio to English.}
\end{tcolorbox}
for ZH$\to$EN, and
\begin{tcolorbox}[colback=gray!5, colframe=black!40, boxrule=0.4pt, arc=2pt]
\footnotesize
\texttt{Translate the above English audio to Chinese. Answer in Chinese only.}
\end{tcolorbox}
for EN$\to$ZH. The explicit target-language constraint is used to reduce language fallback errors.

\paragraph{GPT-Audio.}
For the commercial reference system, we query GPT-Audio with the source clip encoded as audio input and a direction-specific translation instruction:
\begin{tcolorbox}[colback=gray!5, colframe=black!40, boxrule=0.4pt, arc=2pt]
\footnotesize
\texttt{Please translate this Chinese audio into English. Just state the translation itself without any additional explanation or polite words.}
\end{tcolorbox}
for ZH$\to$EN, and
\begin{tcolorbox}[colback=gray!5, colframe=black!40, boxrule=0.4pt, arc=2pt]
\footnotesize
\texttt{Please translate this English audio into Chinese. Just state the translation itself without any additional explanation or polite words.}
\end{tcolorbox}
for EN$\to$ZH. The final clause is included to reduce conversational fillers that would otherwise inflate the generated duration. All other parameters are kept at their default values.

\paragraph{UniSS.}
We evaluate UniSS~\citep{cheng2025uniss} using the official \texttt{cmots/UniSS} checkpoint and inference scripts. UniSS provides two decoding modes, Quality and Performance, which we report as UniSS (Q) and UniSS (P). We use the official target-language tokens for both directions and leave the remaining decoding hyperparameters at their default values.

\subsection{Evaluation Metrics}
\label{app:metric}

We group evaluation metrics into translation quality, voice/prosody preservation, and duration consistency. All scores are computed at the corpus level over the CVSS-T test split.

\paragraph{Translation quality.}
\textbf{Text-BLEU} evaluates the intermediate textual output, while \textbf{Speech-BLEU} evaluates the ASR transcription of the generated speech. Both are computed with SacreBLEU~\citep{post2018sacrebleu} using the \texttt{corpus\_score} interface. English text is lowercased and stripped of punctuation; Chinese text is converted to simplified characters and evaluated in \texttt{zh} mode with character-level tokenization. For Speech-BLEU, we use Whisper-large-v3~\citep{radford2023whisper} for English ASR and Paraformer-zh~\citep{gao2022paraformer} for Mandarin ASR, following Seed-TTS-eval.\footnote{\url{https://github.com/BytedanceSpeech/seed-tts-eval}}

We also report reference-based \textbf{COMET}~\citep{rei2020comet} and reference-free \textbf{COMET\textsubscript{Kiwi}}~\citep{rei2022cometkiwi}, computed with \texttt{Unbabel/wmt22-comet-da} and \texttt{Unbabel/wmt22-cometkiwi-da}, respectively. Both are evaluated on the ASR transcript.

\paragraph{Voice and prosody.}
\textbf{A.PCP}~\citep{barrault2023seamless} measures prosodic consistency between the source and generated speech, including pitch, energy, and temporal patterns. We use the official AutoPCP implementation in \texttt{stopes}.\footnote{\url{https://github.com/facebookresearch/stopes/tree/main/stopes/eval/auto_pcp}} For \textbf{SECS}, we extract speaker embeddings using WavLM-large fine-tuned for speaker verification~\citep{chen2022wavlm}, and compute cosine similarity between the source prompt and generated utterance.

\paragraph{Duration consistency.}
\textbf{SLC-0.2} and \textbf{SLC-0.4}~\citep{wu2023videodubber} measure the fraction of utterances whose target-to-source duration ratio falls within $\pm20\%$ and $\pm40\%$, respectively. Since these threshold-based metrics can saturate under strong duration alignment, we also report two fine-grained error metrics: \textbf{MADE}, the mean absolute duration error in seconds,
\[
\mathrm{MADE} = \mathbb{E}\left[|d_{\mathrm{tgt}}-d_{\mathrm{src}}|\right],
\]
and \textbf{MRDE}, the mean relative duration error,
\[
\mathrm{MRDE} = \mathbb{E}\left[\frac{|d_{\mathrm{tgt}}-d_{\mathrm{src}}|}{d_{\mathrm{src}}}\right].
\]
Both are computed directly from waveform durations.

%% file: sup-sections/eval_example.tex
\section{Evaluation Examples}
\label{app:eval_example}

To complement the quantitative results in Section~\ref{sec:experiments}, we provide qualitative examples from the CVSS-T test set for both translation directions. For each direction, we randomly select two utterances and compare a representative baseline with DuraS2ST. Each example visualizes the synthesized waveform and its duration alignment with the source speech, allowing translation fidelity and temporal synchronization to be inspected jointly.

Figure~\ref{fig:eval_example_en2zh} shows EN$\to$ZH examples, and Figure~\ref{fig:eval_example_zh2en} shows ZH$\to$EN examples. 
For example, in the ZH$\to$EN case, the source utterance describes a company headquartered in Riyadh with branches in Kuwait, Jordan, and Malaysia. 
A compact translation may preserve these facts but still produce target speech that is too short or less naturally paced. 
By reasoning over semantic content and phonetic length before speech-token generation, DuraS2ST better follows the source temporal envelope while producing translations that remain faithful to the reference. 
These examples further demonstrate that the proposed reasoning-based generation and duration-aware optimization generalize to held-out test utterances.

\input{figures/eval_example_en2zh}

\input{figures/eval_example_zh2en}

%% file: figures/eval_example_en2zh.tex
\begin{figure*}[ht]
\centering
\begin{minipage}[t]{0.48\linewidth}
  \centering
  \includegraphics[height=0.50\textheight,keepaspectratio]{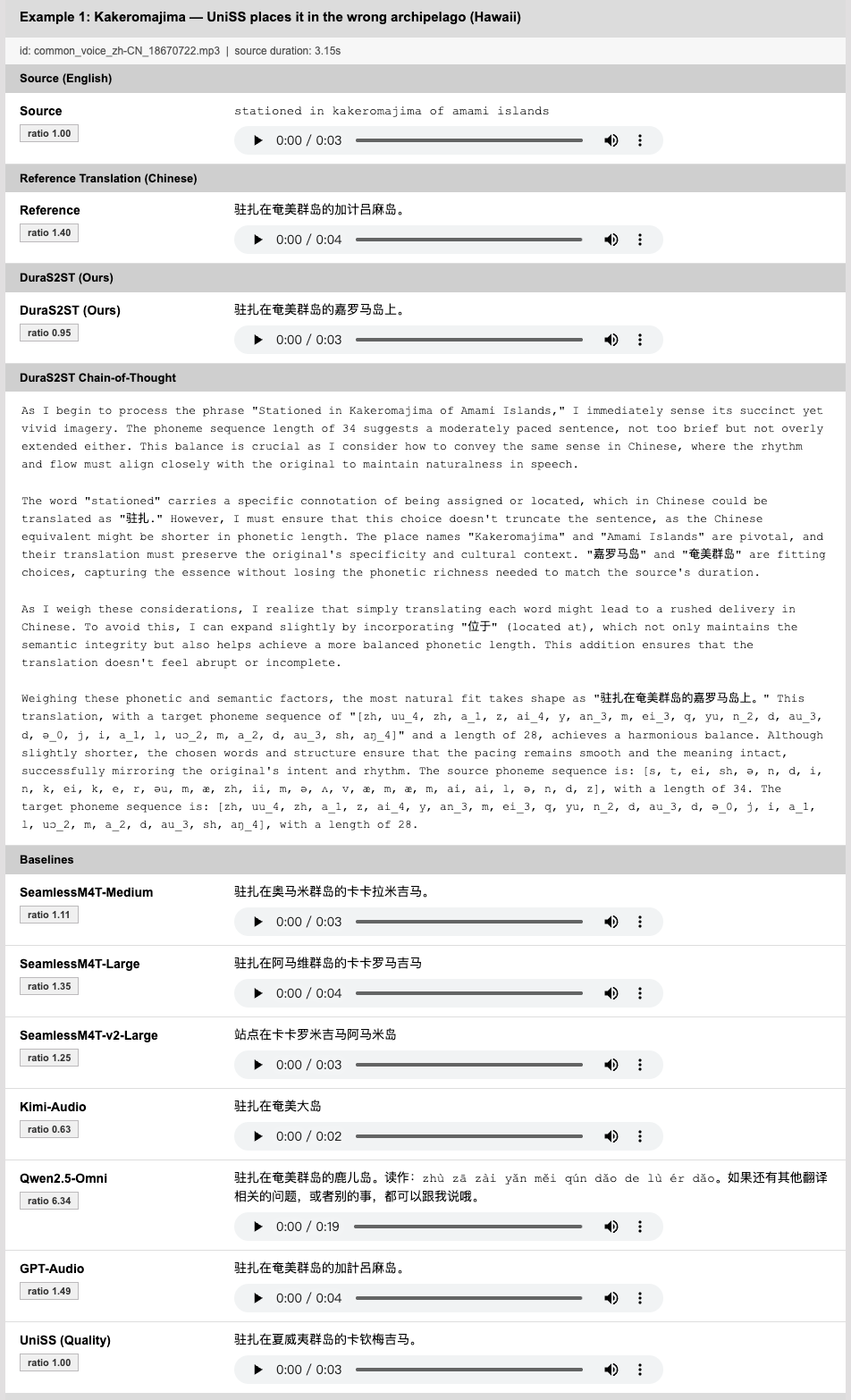}\\[4pt]
  {\small (a) EN$\to$ZH Example~1.}
\end{minipage}\hfill
\begin{minipage}[t]{0.48\linewidth}
  \centering
  \includegraphics[height=0.50\textheight,keepaspectratio]{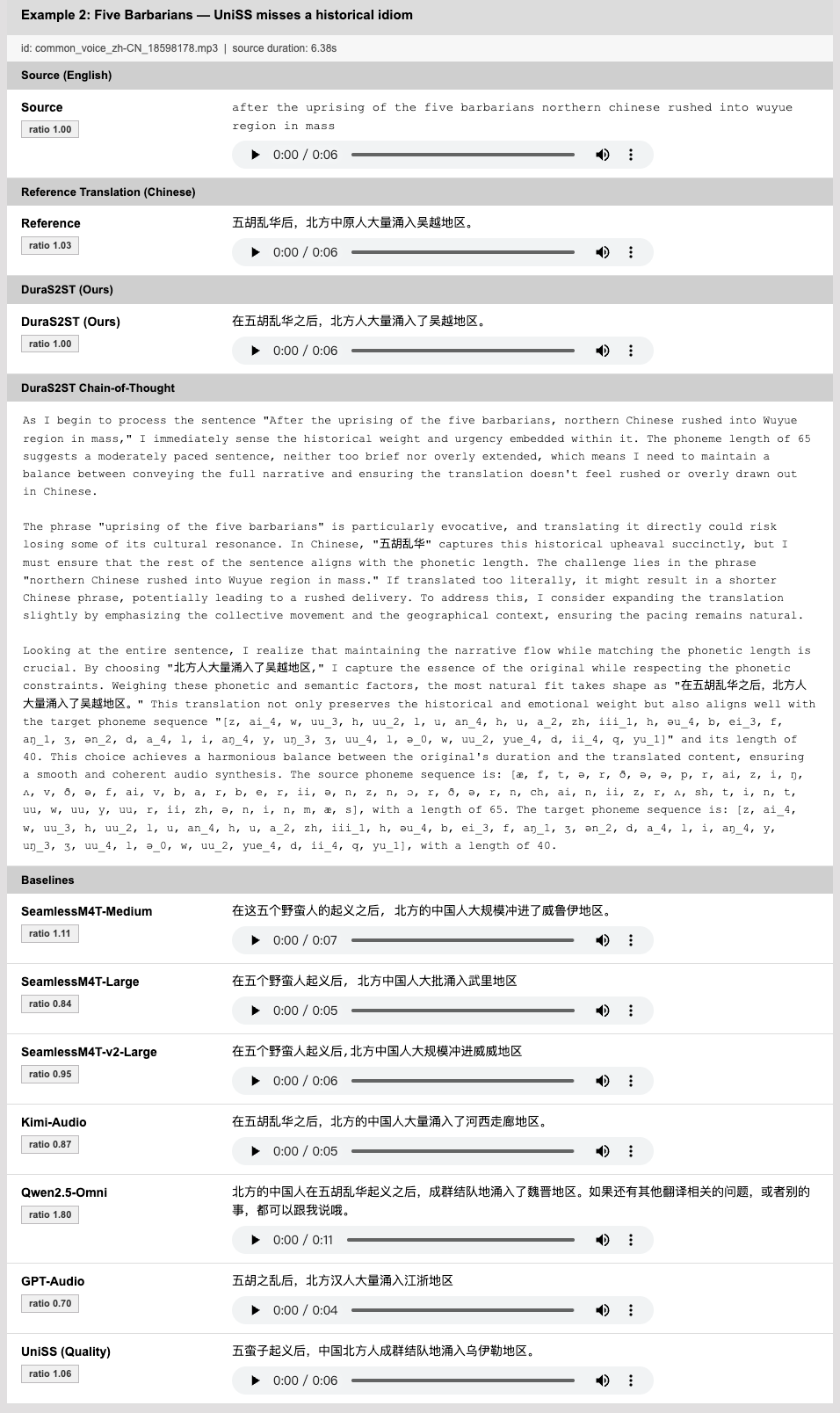}\\[4pt]
  {\small (b) EN$\to$ZH Example~2.}
\end{minipage}
\caption{\textbf{Qualitative EN$\to$ZH test samples.} For each sample we contrast a representative baseline output against the output of DuraS2ST, aligned to the source utterance. The baseline systems exhibit visible duration drift relative to the source's temporal envelope, while DuraS2ST tracks the source duration tightly and produces a translation faithful to the reference.}
\label{fig:eval_example_en2zh}
\end{figure*}

%% file: figures/eval_example_zh2en.tex
\begin{figure*}[ht]
\centering
\begin{minipage}[t]{0.48\linewidth}
  \centering
  \includegraphics[height=0.50\textheight,keepaspectratio]{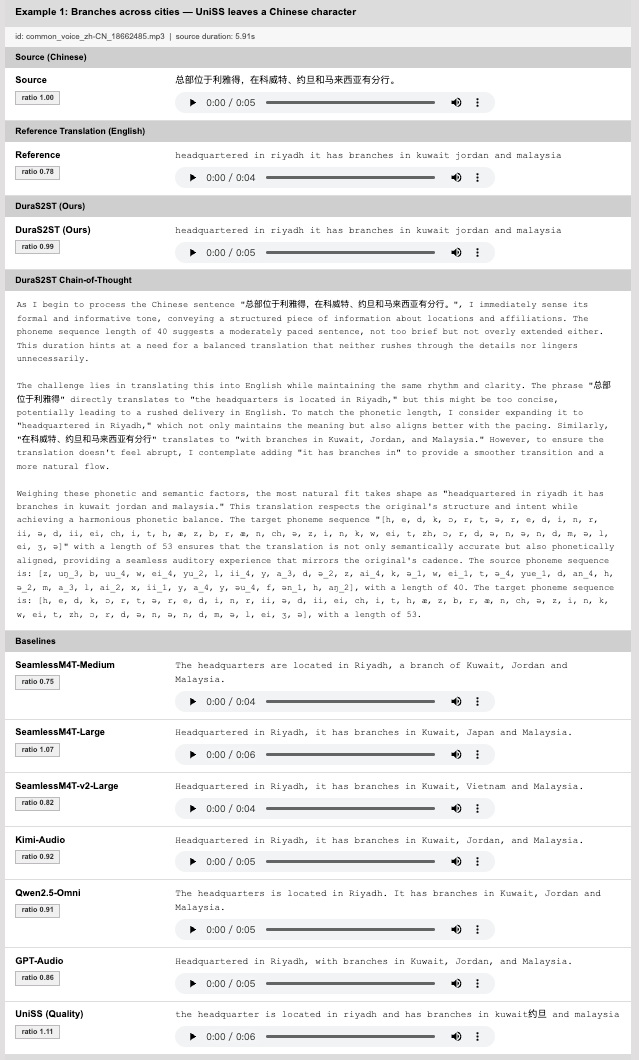}\\[4pt]
  {\small (a) ZH$\to$EN Example~1.}
\end{minipage}\hfill
\begin{minipage}[t]{0.48\linewidth}
  \centering
  \includegraphics[height=0.50\textheight,keepaspectratio]{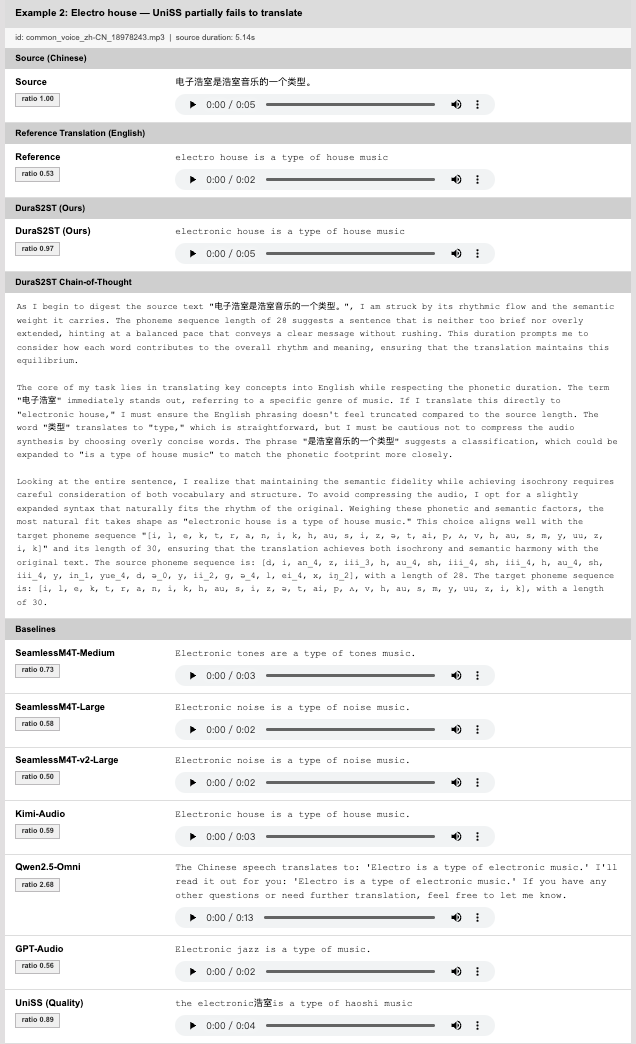}\\[4pt]
  {\small (b) ZH$\to$EN Example~2.}
\end{minipage}
\caption{\textbf{Qualitative ZH$\to$EN test samples.} For each sample we contrast a representative baseline output against the output of DuraS2ST, aligned to the source utterance. As in the EN$\to$ZH case (Figure~\ref{fig:eval_example_en2zh}), DuraS2ST closely matches the source's duration and produces translations consistent with the reference, whereas the baseline drifts in duration and content.}
\label{fig:eval_example_zh2en}
\end{figure*}

%% file: sup-sections/subjective_eval_interface.tex
\section{Subjective Evaluation Interface}
\label{app:mos_interface}

Figure~\ref{fig:subjective_eval_interface} shows the web interface used for the MOS study in Section~\ref{sec:mos}. Each task presents a source clip and one synthesized target clip, with the system identity hidden by an anonymized label. Annotators listen to both clips and provide three integer ratings on a $1$--$5$ Likert scale: \emph{translation adequacy}, measuring whether the target preserves the source meaning; \emph{naturalness}, measuring speech fluency and audio quality; and \emph{speaker similarity}, measuring whether the target preserves the source speaker's voice and prosodic characteristics. The same interface and instructions are used for all $160$ samples across the eight systems. Sample order is randomized for each annotator to reduce presentation bias.

\input{figures/subjective_eval_interface}

%% file: figures/subjective_eval_interface.tex
\begin{figure*}[ht]
\centering
\includegraphics[width=1.0\textwidth,keepaspectratio]{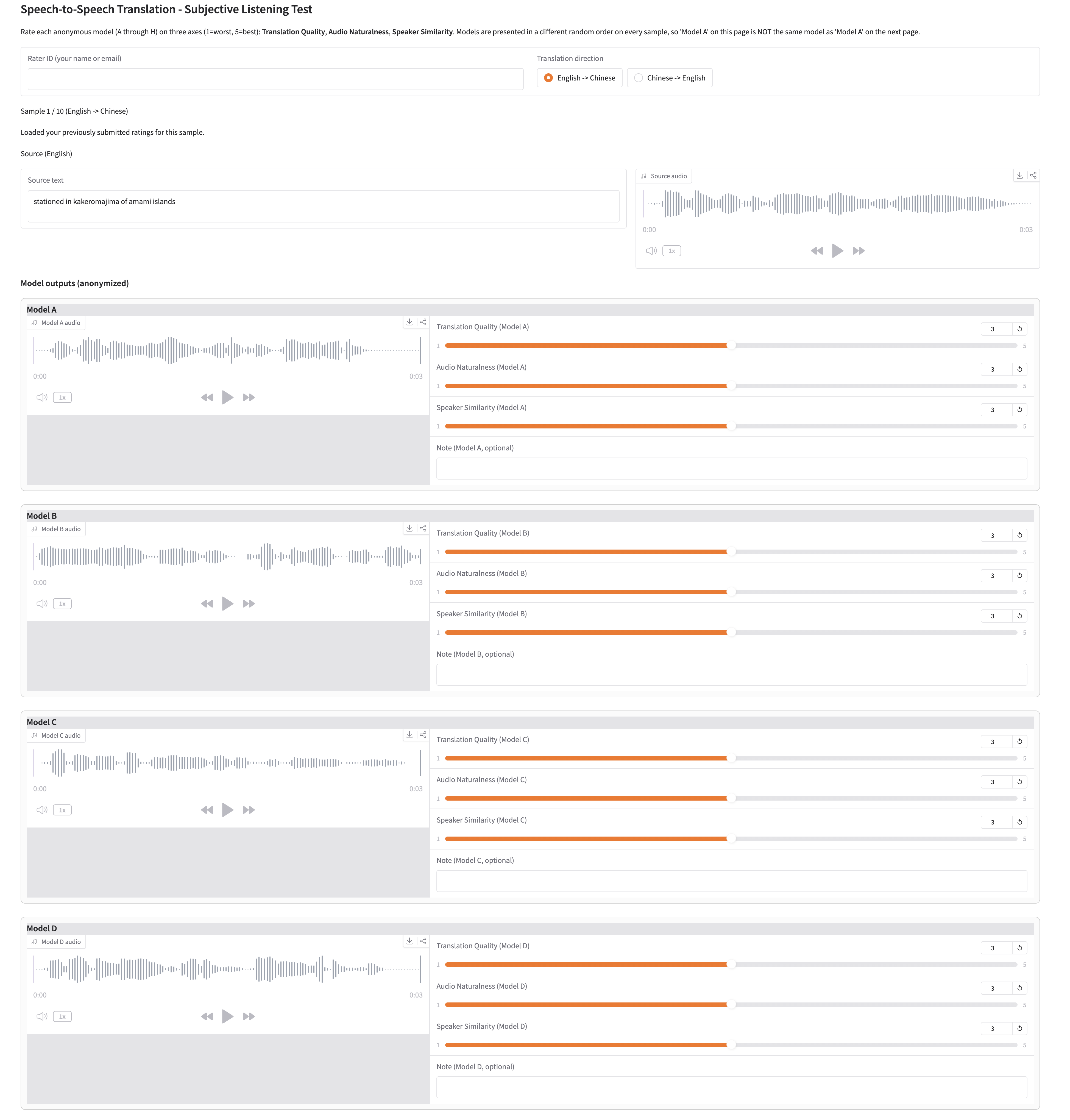}
\caption{\textbf{Subjective evaluation interface used for the MOS study} (Section~\ref{sec:mos}). For every task, the annotator listens to a source clip and a synthesized target clip whose system identity is replaced by an anonymized label, then rates the target on three $1$--$5$ Likert scales---translation adequacy, naturalness, and speaker similarity. The presentation order within each direction is randomized per annotator to mitigate position bias.}
\label{fig:subjective_eval_interface}
\end{figure*}

%% file: acl_latex-dyy.bbl
\begin{thebibliography}{28}
\providecommand{\natexlab}[1]{#1}

\bibitem[{Chen et~al.(2022)Chen, Wang, Chen, Wu, Liu, Chen, Li, Kanda, Yoshioka, Xiao, Wu, Zhou, Ren, Qian, Qian, Wu, Zeng, Yu, and Wei}]{chen2022wavlm}
Sanyuan Chen, Chengyi Wang, Zhengyang Chen, Yu~Wu, Shujie Liu, Zhuo Chen, Jinyu Li, Naoyuki Kanda, Takuya Yoshioka, Xiong Xiao, Jian Wu, Long Zhou, Shuo Ren, Yanmin Qian, Yao Qian, Jian Wu, Michael Zeng, Xiangzhan Yu, and Furu Wei. 2022.
\newblock \href {https://doi.org/10.1109/jstsp.2022.3188113} {Wavlm: Large-scale self-supervised pre-training for full stack speech processing}.
\newblock \emph{IEEE Journal of Selected Topics in Signal Processing}, 16(6):1505–1518.

\bibitem[{Cheng et~al.(2025{\natexlab{a}})Cheng, Bao, Huang, Lu, Peng, Xu, Yu, Cao, Du, Han, Hu, Li, Liu, Ma, Pan, Xiao, Xu, Yang, Ye, Yu, Zhang, Zhang, Zhang, Zhu, Zou, Lu, Wang, and Wu}]{cheng2025seed}
Shanbo Cheng, Yu~Bao, Zhichao Huang, Yu~Lu, Ningxin Peng, Lu~Xu, Runsheng Yu, Rong Cao, Yujiao Du, Ting Han, Yuxiang Hu, Zeyang Li, Sitong Liu, Shengtao Ma, Shiguang Pan, Jiongchen Xiao, Nuo Xu, Meng Yang, Rong Ye, and 9 others. 2025{\natexlab{a}}.
\newblock \href {https://arxiv.org/abs/2507.17527} {Seed liveinterpret 2.0: End-to-end simultaneous speech-to-speech translation with your voice}.
\newblock \emph{Preprint}, arXiv:2507.17527.

\bibitem[{Cheng et~al.(2025{\natexlab{b}})Cheng, Bian, Wang, Yuan, Chen, Yin, Guo, and Xue}]{cheng2025uniss}
Sitong Cheng, Weizhen Bian, Xinsheng Wang, Ruibin Yuan, Jianyi Chen, Shunshun Yin, Yike Guo, and Wei Xue. 2025{\natexlab{b}}.
\newblock \href {https://arxiv.org/abs/2509.21144} {Uniss: Unified expressive speech-to-speech translation with your voice}.
\newblock \emph{Preprint}, arXiv:2509.21144.

\bibitem[{Communication et~al.(2023{\natexlab{a}})Communication, Barrault, Chung, Meglioli, Dale, Dong, Duquenne, Elsahar, Gong, Heffernan, Hoffman, Klaiber, Li, Licht, Maillard, Rakotoarison, Sadagopan, Wenzek, Ye, Akula, Chen, Hachem, Ellis, Gonzalez, Haaheim, Hansanti, Howes, Huang, Hwang, Inaguma, Jain, Kalbassi, Kallet, Kulikov, Lam, Li, Ma, Mavlyutov, Peloquin, Ramadan, Ramakrishnan, Sun, Tran, Tran, Tufanov, Vogeti, Wood, Yang, Yu, Andrews, Balioglu, Costa-jussà, Celebi, Elbayad, Gao, Guzmán, Kao, Lee, Mourachko, Pino, Popuri, Ropers, Saleem, Schwenk, Tomasello, Wang, Wang, and Wang}]{barrault2023seamlessm4t}
Seamless Communication, Loïc Barrault, Yu-An Chung, Mariano~Cora Meglioli, David Dale, Ning Dong, Paul-Ambroise Duquenne, Hady Elsahar, Hongyu Gong, Kevin Heffernan, John Hoffman, Christopher Klaiber, Pengwei Li, Daniel Licht, Jean Maillard, Alice Rakotoarison, Kaushik~Ram Sadagopan, Guillaume Wenzek, Ethan Ye, and 49 others. 2023{\natexlab{a}}.
\newblock \href {https://arxiv.org/abs/2308.11596} {Seamlessm4t: Massively multilingual \& multimodal machine translation}.
\newblock \emph{Preprint}, arXiv:2308.11596.

\bibitem[{Communication et~al.(2023{\natexlab{b}})Communication, Barrault, Chung, Meglioli, Dale, Dong, Duppenthaler, Duquenne, Ellis, Elsahar, Haaheim, Hoffman, Hwang, Inaguma, Klaiber, Kulikov, Li, Licht, Maillard, Mavlyutov, Rakotoarison, Sadagopan, Ramakrishnan, Tran, Wenzek, Yang, Ye, Evtimov, Fernandez, Gao, Hansanti, Kalbassi, Kallet, Kozhevnikov, Gonzalez, Roman, Touret, Wong, Wood, Yu, Andrews, Balioglu, Chen, Costa-jussà, Elbayad, Gong, Guzmán, Heffernan, Jain, Kao, Lee, Ma, Mourachko, Peloquin, Pino, Popuri, Ropers, Saleem, Schwenk, Sun, Tomasello, Wang, Wang, Wang, and Williamson}]{barrault2023seamless}
Seamless Communication, Loïc Barrault, Yu-An Chung, Mariano~Coria Meglioli, David Dale, Ning Dong, Mark Duppenthaler, Paul-Ambroise Duquenne, Brian Ellis, Hady Elsahar, Justin Haaheim, John Hoffman, Min-Jae Hwang, Hirofumi Inaguma, Christopher Klaiber, Ilia Kulikov, Pengwei Li, Daniel Licht, Jean Maillard, and 46 others. 2023{\natexlab{b}}.
\newblock \href {https://arxiv.org/abs/2312.05187} {Seamless: Multilingual expressive and streaming speech translation}.
\newblock \emph{Preprint}, arXiv:2312.05187.

\bibitem[{Dong et~al.(2023)Dong, Huang, Tian, Xu, Ko, Zhao, Feng, Li, Wang, Cheng, Yue, Bai, Chen, Lu, Ma, Wang, Wang, and Wang}]{dong2024polyvoice}
Qianqian Dong, Zhiying Huang, Qiao Tian, Chen Xu, Tom Ko, Yunlong Zhao, Siyuan Feng, Tang Li, Kexin Wang, Xuxin Cheng, Fengpeng Yue, Ye~Bai, Xi~Chen, Lu~Lu, Zejun Ma, Yuping Wang, Mingxuan Wang, and Yuxuan Wang. 2023.
\newblock \href {https://arxiv.org/abs/2306.02982} {Polyvoice: Language models for speech to speech translation}.
\newblock \emph{Preprint}, arXiv:2306.02982.

\bibitem[{Gao et~al.(2023)Gao, Zhang, McLoughlin, and Yan}]{gao2022paraformer}
Zhifu Gao, Shiliang Zhang, Ian McLoughlin, and Zhijie Yan. 2023.
\newblock \href {https://arxiv.org/abs/2206.08317} {Paraformer: Fast and accurate parallel transformer for non-autoregressive end-to-end speech recognition}.
\newblock \emph{Preprint}, arXiv:2206.08317.

\bibitem[{Hu et~al.(2021)Hu, Shen, Wallis, Allen-Zhu, Li, Wang, Wang, and Chen}]{hu2022lora}
Edward~J. Hu, Yelong Shen, Phillip Wallis, Zeyuan Allen-Zhu, Yuanzhi Li, Shean Wang, Lu~Wang, and Weizhu Chen. 2021.
\newblock \href {https://arxiv.org/abs/2106.09685} {Lora: Low-rank adaptation of large language models}.
\newblock \emph{Preprint}, arXiv:2106.09685.

\bibitem[{Huang et~al.(2023)Huang, Liu, Liu, Ren, Zhang, He, and Zhao}]{huang2023transpeech}
Rongjie Huang, Jinglin Liu, Huadai Liu, Yi~Ren, Lichao Zhang, Jinzheng He, and Zhou Zhao. 2023.
\newblock \href {https://arxiv.org/abs/2205.12523} {Transpeech: Speech-to-speech translation with bilateral perturbation}.
\newblock \emph{Preprint}, arXiv:2205.12523.

\bibitem[{Inaguma et~al.(2023)Inaguma, Popuri, Kulikov, Chen, Wang, Chung, Tang, Lee, Watanabe, and Pino}]{inaguma2023unity}
Hirofumi Inaguma, Sravya Popuri, Ilia Kulikov, Peng-Jen Chen, Changhan Wang, Yu-An Chung, Yun Tang, Ann Lee, Shinji Watanabe, and Juan Pino. 2023.
\newblock \href {https://arxiv.org/abs/2212.08055} {Unity: Two-pass direct speech-to-speech translation with discrete units}.
\newblock \emph{Preprint}, arXiv:2212.08055.

\bibitem[{Jia et~al.(2022)Jia, Ramanovich, Wang, and Zen}]{jia2022cvss}
Ye~Jia, Michelle~Tadmor Ramanovich, Quan Wang, and Heiga Zen. 2022.
\newblock \href {https://arxiv.org/abs/2201.03713} {Cvss corpus and massively multilingual speech-to-speech translation}.
\newblock \emph{Preprint}, arXiv:2201.03713.

\bibitem[{KimiTeam et~al.(2025)KimiTeam, Ding, Ju, Leng, Liu, Liu, Shang, Shen, Song, Tan, Tang, Wang, Wei, Xin, Xu, Yu, Zhang, Zhou, Charles, Chen, Chen, Du, He, Hu, Lai, Li, Liu, Sun, Wang, Wang, Wu, Wu, Yang, Yang, Yang, Yang, Yin, Yuan, Zhang, and Zhou}]{ding2025kimi}
KimiTeam, Ding Ding, Zeqian Ju, Yichong Leng, Songxiang Liu, Tong Liu, Zeyu Shang, Kai Shen, Wei Song, Xu~Tan, Heyi Tang, Zhengtao Wang, Chu Wei, Yifei Xin, Xinran Xu, Jianwei Yu, Yutao Zhang, Xinyu Zhou, Y.~Charles, and 21 others. 2025.
\newblock \href {https://arxiv.org/abs/2504.18425} {Kimi-audio technical report}.
\newblock \emph{Preprint}, arXiv:2504.18425.

\bibitem[{Kwon et~al.(2023)Kwon, Li, Zhuang, Sheng, Zheng, Yu, Gonzalez, Zhang, and Stoica}]{kwon2023efficient}
Woosuk Kwon, Zhuohan Li, Siyuan Zhuang, Ying Sheng, Lianmin Zheng, Cody~Hao Yu, Joseph~E. Gonzalez, Hao Zhang, and Ion Stoica. 2023.
\newblock \href {https://arxiv.org/abs/2309.06180} {Efficient memory management for large language model serving with pagedattention}.
\newblock \emph{Preprint}, arXiv:2309.06180.

\bibitem[{Le et~al.(2024)Le, Qian, Wang, Zhou, Liu, Wang, Yousefi, Qian, Li, Zhao, and Zeng}]{le2024transvip}
Chenyang Le, Yao Qian, Dongmei Wang, Long Zhou, Shujie Liu, Xiaofei Wang, Midia Yousefi, Yanmin Qian, Jinyu Li, Sheng Zhao, and Michael Zeng. 2024.
\newblock \href {https://arxiv.org/abs/2405.17809} {Transvip: Speech to speech translation system with voice and isochrony preservation}.
\newblock \emph{Preprint}, arXiv:2405.17809.

\bibitem[{Lee et~al.(2022)Lee, Chen, Wang, Gu, Popuri, Ma, Polyak, Adi, He, Tang, Pino, and Hsu}]{lee2022direct}
Ann Lee, Peng-Jen Chen, Changhan Wang, Jiatao Gu, Sravya Popuri, Xutai Ma, Adam Polyak, Yossi Adi, Qing He, Yun Tang, Juan Pino, and Wei-Ning Hsu. 2022.
\newblock \href {https://arxiv.org/abs/2107.05604} {Direct speech-to-speech translation with discrete units}.
\newblock \emph{Preprint}, arXiv:2107.05604.

\bibitem[{Post(2018)}]{post2018sacrebleu}
Matt Post. 2018.
\newblock \href {https://arxiv.org/abs/1804.08771} {A call for clarity in reporting bleu scores}.
\newblock \emph{Preprint}, arXiv:1804.08771.

\bibitem[{Radford et~al.(2022)Radford, Kim, Xu, Brockman, McLeavey, and Sutskever}]{radford2023whisper}
Alec Radford, Jong~Wook Kim, Tao Xu, Greg Brockman, Christine McLeavey, and Ilya Sutskever. 2022.
\newblock \href {https://arxiv.org/abs/2212.04356} {Robust speech recognition via large-scale weak supervision}.
\newblock \emph{Preprint}, arXiv:2212.04356.

\bibitem[{Rei et~al.(2020)Rei, Stewart, Farinha, and Lavie}]{rei2020comet}
Ricardo Rei, Craig Stewart, Ana~C Farinha, and Alon Lavie. 2020.
\newblock \href {https://doi.org/10.18653/v1/2020.emnlp-main.213} {{COMET}: A neural framework for {MT} evaluation}.
\newblock In \emph{Proceedings of the 2020 Conference on Empirical Methods in Natural Language Processing (EMNLP)}, pages 2685--2702, Online. Association for Computational Linguistics.

\bibitem[{Rei et~al.(2022)Rei, Treviso, Guerreiro, Zerva, Farinha, Maroti, C.~de Souza, Glushkova, Alves, Coheur, Lavie, and Martins}]{rei2022cometkiwi}
Ricardo Rei, Marcos Treviso, Nuno~M. Guerreiro, Chrysoula Zerva, Ana~C Farinha, Christine Maroti, Jos{\'e}~G. C.~de Souza, Taisiya Glushkova, Duarte Alves, Luisa Coheur, Alon Lavie, and Andr{\'e} F.~T. Martins. 2022.
\newblock \href {https://doi.org/10.18653/v1/2022.wmt-1.60} {{C}omet{K}iwi: {IST}-unbabel 2022 submission for the quality estimation shared task}.
\newblock In \emph{Proceedings of the Seventh Conference on Machine Translation (WMT)}, pages 634--645, Abu Dhabi, United Arab Emirates (Hybrid). Association for Computational Linguistics.

\bibitem[{Rubenstein et~al.(2023)Rubenstein, Asawaroengchai, Nguyen, Bapna, Borsos, de~Chaumont~Quitry, Chen, Badawy, Han, Kharitonov, Muckenhirn, Padfield, Qin, Rozenberg, Sainath, Schalkwyk, Sharifi, Ramanovich, Tagliasacchi, Tudor, Velimirović, Vincent, Yu, Wang, Zayats, Zeghidour, Zhang, Zhang, Zilka, and Frank}]{rubenstein2023audiopalm}
Paul~K. Rubenstein, Chulayuth Asawaroengchai, Duc~Dung Nguyen, Ankur Bapna, Zalán Borsos, Félix de~Chaumont~Quitry, Peter Chen, Dalia~El Badawy, Wei Han, Eugene Kharitonov, Hannah Muckenhirn, Dirk Padfield, James Qin, Danny Rozenberg, Tara Sainath, Johan Schalkwyk, Matt Sharifi, Michelle~Tadmor Ramanovich, Marco Tagliasacchi, and 11 others. 2023.
\newblock \href {https://arxiv.org/abs/2306.12925} {Audiopalm: A large language model that can speak and listen}.
\newblock \emph{Preprint}, arXiv:2306.12925.

\bibitem[{Shao et~al.(2024)Shao, Wang, Zhu, Xu, Song, Bi, Zhang, Zhang, Li, Wu, and Guo}]{shao2024deepseekmath}
Zhihong Shao, Peiyi Wang, Qihao Zhu, Runxin Xu, Junxiao Song, Xiao Bi, Haowei Zhang, Mingchuan Zhang, Y.~K. Li, Y.~Wu, and Daya Guo. 2024.
\newblock \href {https://arxiv.org/abs/2402.03300} {Deepseekmath: Pushing the limits of mathematical reasoning in open language models}.
\newblock \emph{Preprint}, arXiv:2402.03300.

\bibitem[{Wahlster(2000)}]{wahlsterverbmobil}
Wolfgang Wahlster. 2000.
\newblock \emph{Verbmobil: foundations of speech-to-speech translation}, volume~1.
\newblock Springer Berlin.

\bibitem[{Wei et~al.(2022)Wei, Bosma, Zhao, Guu, Yu, Lester, Du, Dai, and Le}]{wei2022finetuned}
Jason Wei, Maarten Bosma, Vincent~Y. Zhao, Kelvin Guu, Adams~Wei Yu, Brian Lester, Nan Du, Andrew~M. Dai, and Quoc~V. Le. 2022.
\newblock \href {https://arxiv.org/abs/2109.01652} {Finetuned language models are zero-shot learners}.
\newblock \emph{Preprint}, arXiv:2109.01652.

\bibitem[{Wu et~al.(2025)Wu, Yan, Hu, Yi, Feng, Tian, Shen, Yu, Zhang, Li, Chen, Liu, You, Zhang, Li, Yang, Deng, Huang, Li, Zhang, You, Li, Wan, Hu, Zhen, Chen, Yuan, Zhang, Jiang, Zhou, Yang, Li, Ma, Song, Pang, Hu, Sun, An, Wang, Gao, Ji, Li, Sun, Wen, Ren, Ma, Lu, Wang, Li, Miao, Liu, Xu, Shi, Hu, Wu, Liu, Huang, Yan, Zhang, Nie, Jia, Zhou, Sun, Wu, Wu, Yang, Yang, Lin, Li, Yang, Shi, Zhou, Gu, Li, Li, Li, Wu, Han, Tan, Pang, Fan, Liu, Cao, Lu, He, Xie, Zhao, Li, Yu, Yang, Liu, Lu, Wang, Ding, Liang, Lu, Luo, Yin, Zhan, Zhang, Yang, Zhang, Jiao, Jiang, Shum, Chen, Li, Zhang, and Zhu}]{wu2025step}
Boyong Wu, Chao Yan, Chen Hu, Cheng Yi, Chengli Feng, Fei Tian, Feiyu Shen, Gang Yu, Haoyang Zhang, Jingbei Li, Mingrui Chen, Peng Liu, Wang You, Xiangyu~Tony Zhang, Xingyuan Li, Xuerui Yang, Yayue Deng, Yechang Huang, Yuxin Li, and 90 others. 2025.
\newblock \href {https://arxiv.org/abs/2507.16632} {Step-audio 2 technical report}.
\newblock \emph{Preprint}, arXiv:2507.16632.

\bibitem[{Wu et~al.(2023)Wu, Guo, Tan, Zhang, Li, Song, He, Zhao, Menezes, and Bian}]{wu2023videodubber}
Yihan Wu, Junliang Guo, Xu~Tan, Chen Zhang, Bohan Li, Ruihua Song, Lei He, Sheng Zhao, Arul Menezes, and Jiang Bian. 2023.
\newblock \href {https://arxiv.org/abs/2211.16934} {Videodubber: Machine translation with speech-aware length control for video dubbing}.
\newblock \emph{Preprint}, arXiv:2211.16934.

\bibitem[{Xu et~al.(2025)Xu, Guo, He, Hu, He, Bai, Chen, Wang, Fan, Dang, Zhang, Wang, Chu, and Lin}]{xu2025qwen25omni}
Jin Xu, Zhifang Guo, Jinzheng He, Hangrui Hu, Ting He, Shuai Bai, Keqin Chen, Jialin Wang, Yang Fan, Kai Dang, Bin Zhang, Xiong Wang, Yunfei Chu, and Junyang Lin. 2025.
\newblock \href {https://arxiv.org/abs/2503.20215} {Qwen2.5-omni technical report}.
\newblock \emph{Preprint}, arXiv:2503.20215.

\bibitem[{Yu et~al.(2025)Yu, Zhang, Zhu, Yuan, Zuo, Yue, Dai, Fan, Liu, Liu, Liu, Lin, Lin, Ma, Sheng, Tong, Zhang, Zhang, Zhang, Zhu, Zhu, Chen, Chen, Wang, Yu, Song, Wei, Zhou, Liu, Ma, Zhang, Yan, Qiao, Wu, and Wang}]{yu2026dapo}
Qiying Yu, Zheng Zhang, Ruofei Zhu, Yufeng Yuan, Xiaochen Zuo, Yu~Yue, Weinan Dai, Tiantian Fan, Gaohong Liu, Lingjun Liu, Xin Liu, Haibin Lin, Zhiqi Lin, Bole Ma, Guangming Sheng, Yuxuan Tong, Chi Zhang, Mofan Zhang, Wang Zhang, and 16 others. 2025.
\newblock \href {https://arxiv.org/abs/2503.14476} {Dapo: An open-source llm reinforcement learning system at scale}.
\newblock \emph{Preprint}, arXiv:2503.14476.

\bibitem[{Zhao et~al.(2026)Zhao, Lin, Zhu, Xie, Liu, and Li}]{zhao2026lemas}
Zhiyuan Zhao, Lijian Lin, Ye~Zhu, Kai Xie, Yunfei Liu, and Yu~Li. 2026.
\newblock \href {https://arxiv.org/abs/2601.04233} {Lemas: Large a 150k-hour large-scale extensible multilingual audio suite with generative speech models}.
\newblock \emph{Preprint}, arXiv:2601.04233.

\end{thebibliography}
